\documentclass[11pt, a4paper]{article}
\usepackage[margin=1in]{geometry}
\usepackage{graphicx} % Required for inserting images
\usepackage[round]{natbib}
\usepackage{listings}
\usepackage{color}
\usepackage{subcaption}
\usepackage{hyperref}
\usepackage{amsmath}
\usepackage{amssymb}
\usepackage{booktabs}

\newtheorem{definition}{Definition}
\newtheorem{example}{Example}
\newtheorem{theorem}{Theorem}
\newtheorem{corollary}{Corollary}

\title{Balancing the privacy-utility trade-off: How to draw reliable conclusions from private data}

\author{Raphaël de Fondeville \\ Federal Statistical Office, Neuchâtel, Switzerland}

\begin{document}
% At least three keywords are required at submission. Please provide three to five keywords, separated by the pipe symbol.
%\keywords{Differential Privacy $|$ Hypothesis Testing $|$ Membership Attack $|$ Relative Risk $|$ Statistical Inference.}
\maketitle

\begin{abstract}

Absolute anonymization, conceived as an irreversible transformation preventing re-identi-fication and sensitive value disclosure, has proven to be a broken promise. Modern data protection must therefore shift toward a privacy-utility trade-off grounded in risk mitigation. Differential Privacy (DP) offers a rigorous mathematical framework for balancing quantified disclosure risk with analytical usefulness. Nevertheless, widespread adoption remains limited, largely because complex technical concepts, such as privacy-loss parameters, have yet to be translated into forms meaningful to non-technical stakeholders.
This difficulty arises from randomization itself: both analysts and adversaries must draw conclusions from uncertain observations rather than deterministic values.
In this work, we adopt an interpretation of the privacy-utility trade-off based on hypothesis testing to measure the uncertainty introduced by randomized mechanisms. In particular, we use the concept of relative disclosure risk to quantify the maximum reduction in uncertainty an adversary can obtain from a membership attack on protected outputs, and show this measure relates directly to standard privacy-loss parameters. We further analyze how DP affects analytical validity via its impact on hypothesis tests assessing statistical significance.
Building on these results, we provide practical guidance, accessible to non-experts such as data protection authorities, for navigating the trade-off and selecting protection mechanisms and parameter values.
\end{abstract}

\section{Introduction}
Privacy protection has long been intertwined with technological progress. What is considered “acceptable” when it comes to privacy is deeply shaped by social norms, political values, and historical context. A striking example is the rise of portable photography in the late 19th century, which allowed people to capture candid images without consent. At the time, this was seen by many as a serious threat, some even feared it marked the end of privacy altogether. The resulting public backlash led to the landmark 1890 article by Warren and Brandeis \citep{Warren1980}, which laid the foundation for the legal concept of the right to privacy in the United States. That reaction may seem surprising today, given how widely accepted and commonplace personal photography has become.

Today, a new wave of technological change, driven by digitalization and the widespread systematic collection of personal data, poses fresh challenges to privacy. In response, many efforts have focused on anonymization: transforming data so that individuals are no longer identifiable. But the idea of absolute anonymization, i.e., completely irreversible transformations, has proven unrealistic. Traditional techniques like aggregation, thresholding, and $k$-anonymity \citep{sweeney2002} have all been shown to be vulnerable to increasingly sophisticated attacks such as linkage \cite[e.g.,][]{sweeney2000}, reconstruction \cite[e.g.,][]{abowd2023}, and membership inference \cite[e.g.,][]{homer2008}.
Given the limits of these approaches, it's clear that privacy protection must evolve. Rather than aiming for a binary notion of “anonymous or not,” the focus should shift toward managing the trade-off between two inherently conflicting goals: maximizing the usefulness of data and minimizing the risk of disclosure.

%Among existing approaches, randomization techniques—and in particular the rigorous mathematical framework of differential privacy (DP) \citep{dwork2006}—are now widely regarded as the most promising solution for managing disclosure risk when disseminating algorithms results, also called data products, generated or  derived data.
Among existing approaches, randomization techniques, particularly the rigorous mathematical framework of differential privacy (DP) \citep{dwork2006}, are widely regarded as the most promising solution for managing disclosure risk in the dissemination of algorithmic outputs.
DP is especially compelling due to properties such as ``composition'', which enables automatic cumulative privacy accounting across multiple algorithms of varying nature; and ``post-processing'', which ensures that once results have been protected under DP, any further processing cannot degrade the privacy guarantee, making it robust to future technological advances and increasing availability of auxiliary information.
Nevertheless, the adoption of DP faces several challenges. Chief among them are the conceptual complexity of its guarantees, the difficulty of selecting appropriate privacy-loss parameters, and the challenge of communicating their implications to non-technical stakeholders. While DP offers precise, mathematically defined control over disclosure risk, deciding what constitutes an “acceptable” risk level remains a fundamentally socio-political and context-dependent decision that must account for the nature of the data, the expected benefits of data use, and the recipients of the disseminated results.
%This challenge is well illustrated by the impact of portable photography in the late 19th century, which enabled unauthorized and candid image capture, provoked public outrage, and ultimately led to the landmark 1890 article by Warren and Brandeis \citep{Warren1980} that laid the legal foundation for the right to privacy in the United States—a stark contrast to the widespread acceptance of such practices in today's society.
Thus, a broader adoption of differential privacy necessitates interpretable, tangible representations of the guarantees it provides, expressed in terms that are intelligible to policy-makers, legal experts, regulators, and other non-technical decision-makers.

The notion of privacy under scrutiny throughout this paper is the confidentiality of data that has been collected, not the broader risk that sensitive information about an individual could be inferred independently of the presence of their data in the dataset, a distinction that mirrors data protection regulations such as the GDPR or the Swiss Federal Act on Data Protection (LPD), which govern the processing of collected personal data.
In this context, we propose
%a novel and practical interpretation of differential privacy guarantees by leveraging hypothesis testing and the $f$-DP framework \citep{Dong2022}. Our approach provides
an interpretable framework for making the impact of randomization on both legitimate analysts and potential adversaries explicit and interpretable.
%the reliability of membership attacks, i.e., data disclosure by assessment of an individual's membership to a group of the population, and the impact of randomization on statistical inference can both be assessed.
After reviewing the fundamental concepts of differential privacy and hypothesis testing in Section \ref{sec: background}, Section \ref{sec: privacy garantee} analyzes the performance of membership attacks, i.e., attacks aimed at determining whether a specific individual’s data is included in a dataset, when conducted on DP-protected results.
By randomizing algorithmic outputs, differential privacy converts a known deterministic vulnerability,
i.e., the existence of a membership attacks scheme revealing with certainty whether an individual’s data record is included in a given dataset, into a quantifiable and controllable risk.
%Our framework offers a new perspective on DP’s core guarantee, framing it as a mechanism that prevents definitive conclusions and compels adversaries to operate under uncertainty.
%The challenge of making reliable conclusions under uncertainty has long been addressed in statistics, particularly through the framework of hypothesis testing.
%Leveraging the formalism offered by statistical hypothesis tests, we offer a principled characterization of the unreliability of membership attacks in the context of DP.
Building on the Bayesian reading of differential privacy established by \cite{kasiviswanathan2014semantics} and the hypothesis-testing formulation of \cite{wasserman2010statistical}, our framework does not introduce a new mathematical guarantee, but offers a unified and interpretable perspective on DP's core guarantee:: framed through the hypothesis-testing formalism of \cite{Dong2022}, DP prevents definitive conclusions and compels adversaries to operate under uncertainty.
Leveraging this existing formalism, we offer a principled characterization of the unreliability of membership attacks in the context of DP, assembling quantities already implicit in the differential privacy literature into a coherent, decision-relevant picture.
%Specifically, we identify the conditions under which blatant disclosure may occur and directly link the privacy-loss parameters to concrete statistical measures, such as the statistical power of an attack.
Specifically, building on informal characterizations from prior work directed toward machine learning applications \citep{ponomareva2023dpfy} and a formal characterization by \cite{kulynych2024attackaware}, we identify the conditions under which blatant disclosure may occur and directly link the privacy-loss parameters to concrete statistical measures, such as the statistical power of an attack.
%We introduce the concept of relative disclosure risk, which quantifies the maximum factor by which an adversary’s prior belief can be updated after observing a DP-protected result.
We adopt the concept of relative disclosure risk, originally proposed by \cite{mcclure2012differential} and used by \cite{kazan2023prioritizing} to guide the choice of privacy-loss budget, but in a form designed for interpretability, as a property of the mechanism itself rather than of any particular adversary. It quantifies the maximum factor by which an adversary's prior belief can be updated after observing a DP-protected result, independently of the attacker's knowledge.
%{Framing privacy loss as a bound on how much beliefs can change makes the guarantee more transparent to non-experts and practitioners, and we view this interpretability as a central contribution of our discussion, especially given that relative risk has proven to be a successful and widely understood metric in other domains, such as epidemiology.}
Framing privacy loss as a bound on how much beliefs can change, independent of any particular attacker, makes the guarantee more transparent to non-experts and practitioners, especially given that relative risk has proven to be a successful and widely understood metric in other domains, such as epidemiology.

%and frames membership attacks as a hypothesis testing problem.
%This enables us to rigorously characterize the unreliability of membership attacks. Specifically, we examine the conditions under which blatant disclosure occurs and relate the privacy-loss parameters directly to metrics such as relative disclosure risk and attack power.

Section \ref{sec: utility quantification} adopts the perspective of the legitimate analyst: like the hypothetical attacker, the analyst is affected by the randomization and must account for this additional source of uncertainty.
Hypothesis testing has historically provided a rigorous framework for drawing conclusions from a finite set of observations, which are often subject to inherent uncertainty. In this context, DP can be viewed simply as an additional, well-characterized source of uncertainty.
%Accounting for DP is therefore conceptually equivalent to adjusting standard statistical analyses to reflect this new quantifiable variability, and thus allows analysts to draw valid inferences while preserving privacy.
We leverage our framework to explain how the introduction of a DP mechanism influences standard statistical analyses.
We specifically quantify the impact of DP on the statistical power of hypothesis tests, providing a concrete and interpretable metric that links the degradation of utility from privacy constraints directly to an analyst’s ability to draw reliable conclusions.
%The methodology must be adapted to each specific analysis, according to the nature of the measurements’ uncertainty. For illustrative purposes, we apply this procedure to a commonly used test of mean differences under Gaussian assumptions.

%From a utility perspective, Section 3 relies once again on hypothesis testing to assess how the randomized mechanism impacts statistical analyses—particularly for research application in domains like public health or economics. We quantify how DP affects statistical power, thus providing an interpretable measure of utility degradation under privacy constraints.

Finally, Section \ref{sec: trade-off} presents the privacy–utility trade-off and the two paradigms that facilitate its management: the privacy-first and utility-first strategies.
While the former is the default mode of thinking in most applications of differential privacy, the latter is used more rarely, despite having been studied in dedicated contexts, notably by the U.S. Census Bureau in its demonstration data products \citep{abowd2022topdown} and in optimization frameworks tailored to specific analytical tasks \citep{ligett2017accuracy, whitehouse2022brownian}.
These paradigms enable the balancing of disclosure risk against expected benefits while accounting for the nature of the data, the intended use of the results, and their recipient(s).
To support practical application, we provide guidelines for selecting privacy-loss parameters under both strategies, structured as a set of questions designed to facilitate decision-making. We then discuss their implications, connecting the choice of a differential privacy mechanism and its parameters to concrete considerations for stakeholders without specialized expertise in privacy technologies.

\section{Materials and Methods}\label{sec: background}
%\matmethods{
%We start by reviewing essential background on (Bayesian) hypothesis testing and the $f$-DP framework.
\subsection{Differential Privacy}\label{sec: dp}

Differential privacy \citep{dwork2006} was originally introduced to constrain an adversary’s ability to distinguish between neighboring datasets, i.e., any pair of datasets $D$ and $D'$ differing by exactly one privacy unit, based on the output of a randomized mechanism.
\begin{definition}
    A {\bf Randomized Mechanism} is an algorithm that takes as input a data set $D$ and outputs a randomized version $Q(D)$ of an algorithm $q(D)$, also known as a query.
\end{definition}
The privacy unit refers to the entity whose privacy is being protected. While typically representing an individual, physical or legal, it may also correspond to other granularities such as households, records, or multiple entries associated with a single person, depending on the context. The privacy guarantee ensures that the inclusion or exclusion of this unit in the dataset has a controlled effect on the mechanism's output, thereby preserving the confidentiality of their data.

A randomized mechanism $Q(D)$ satisfies differential privacy with a privacy-loss budget $\epsilon > 0$ if, for any neighboring datasets $D$ and $D'$, and for any event $E$, \begin{equation}\label{eq: pure dp} \frac{\text{Pr}\{Q(D') \in E\}}{\text{Pr}\{Q(D) \in E\}} \leq e^\epsilon, \end{equation} where $E$ belongs to the Borel sigma-algebra over the range of possible outputs of $Q(D)$.
This formulation, introduced by \cite{dwork2006}, is known as pure differential privacy. Despite its strong theoretical guarantees, its practical adoption has often been hindered by the challenge of selecting a meaningful value for the privacy-loss parameter $\epsilon$, especially for audiences lacking a mathematical background.

One of the most effective conceptual tools for interpreting $\epsilon$ involves counterfactual reasoning: the privacy budget regulates the distinguishability between two scenarios, one in which an individual participates in the dataset, and one in which they do not. In this view, $\epsilon$ controls the adversary's ability to infer an individual’s participation, thereby quantifying the potential harm associated with their inclusion.
This counterfactual perspective, commonly known as posterior-to-posterior analysis \citep{Kifer2022}, effectively conveys the foundational idea behind the DP guarantee. However, it falls short of offering a concrete and operational interpretation of the parameter $\epsilon$ and its possible values.

Efforts to guide the selection of privacy budgets have primarily relied on cataloging initiatives \cite[e.g.,][]{Dwork2019}, which list examples of $\epsilon$-values used across various DP-protected data releases. Although useful for reference, these catalogs are difficult to generalize: the appropriate choice of $\epsilon$ is highly context-dependent, influenced by factors such as legal and regulatory constraints, geographic jurisdiction, temporal considerations, data sensitivity, query structure, and the scope of the data recipients, i.e., the group of individuals with access to the algorithm’s outputs.

\subsection{Hypothesis Testing and $f$-Differential Privacy}\label{sec: f-dp}
Building upon \citep{kairouz2015composition}, a more general framework known as $f$-Differential Privacy ($f$-DP) has recently been introduced \citep{Dong2022}, and refined \citep{Su2025}, offering a deeper and more comprehensive understanding of the guarantees provided by randomized mechanisms. Rather than focusing solely on the inequality in \eqref{eq: pure dp}, $f$-DP aims to restrict an adversary’s ability to infer sensitive information by limiting their capacity to distinguish, for any pair of neighboring datasets $D$ and $D'$, between the hypotheses:
\begin{equation}\label{eq: hyp f-dp}
    \begin{array}{ll}
        H_0: & \text{the randomized mechanism used } D \text{ as input}, \\
        H_1: & \text{the randomized mechanism used } D' \text{ as input}.
    \end{array}
\end{equation}
The distinction between $H_0$ and $H_1$ is evaluated using a statistical hypothesis test. Given an observed output $q$, a univariate test statistic $T \in \mathbb{R}$ is computed via a function derived from the distributions of $Q(D)$ and $Q(D')$. If the statistic exceeds a certain threshold $\phi \in \mathbb{R}$, the test rejects $H_0$ in favor of $H_1$ with a given level of confidence.

The choice of threshold $\phi$ defines the following error rates:
\begin{itemize}
\item \textbf{Type I error} ($\alpha = \alpha(\phi)$): the probability of incorrectly rejecting $H_0$ when the output $q$ was actually generated from $Q(D)$; also known as a false positive.
\item \textbf{Type II error} ($\beta = \beta(\phi)$): the probability of failing to reject $H_0$ when $q$ was generated from $Q(D')$; also known as a false negative.
\end{itemize}
The relationship between $\alpha$ and $\beta$ fully characterizes the ability to distinguish between the two hypotheses. Perfect distinguishability would require both errors to be simultaneously zero, which is generally impossible. Therefore, the threshold $\phi$ must be selected to balance these errors, based on the contextual consequences of each type of misclassification:
\begin{itemize}
\item A small $\alpha$ is preferred when false positives are particularly harmful.
\item A small $\beta$ is favored when false negatives carry greater risk.
\end{itemize}
Because the optimal trade-off depends heavily on context, it is often more informative to study the entire spectrum of possible $(\alpha,\beta)$ pairs. This relationship is commonly visualized via Receiver Operating Characteristic (ROC) curves \citep{Marcum1947}, which plot the test power $1 - \beta$ as a function of $\alpha$. ROC curves are widely used not only in hypothesis testing but also in machine learning for instance to evaluate classifier performance.

\begin{figure}%[tbhp]
\centering
\begin{tabular}{ccc}
     \includegraphics[width=.28\linewidth]{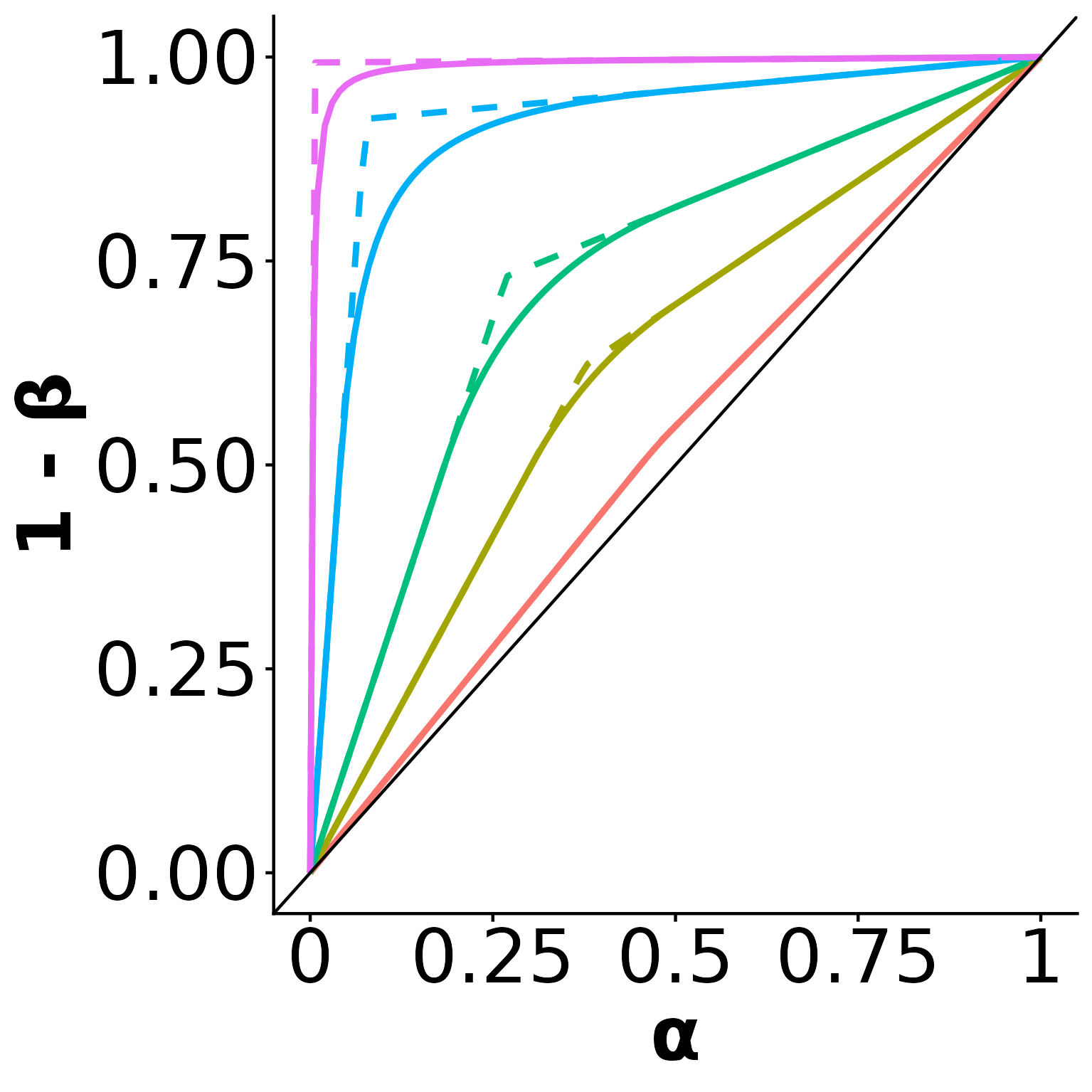} &
     \includegraphics[width=.28\linewidth]{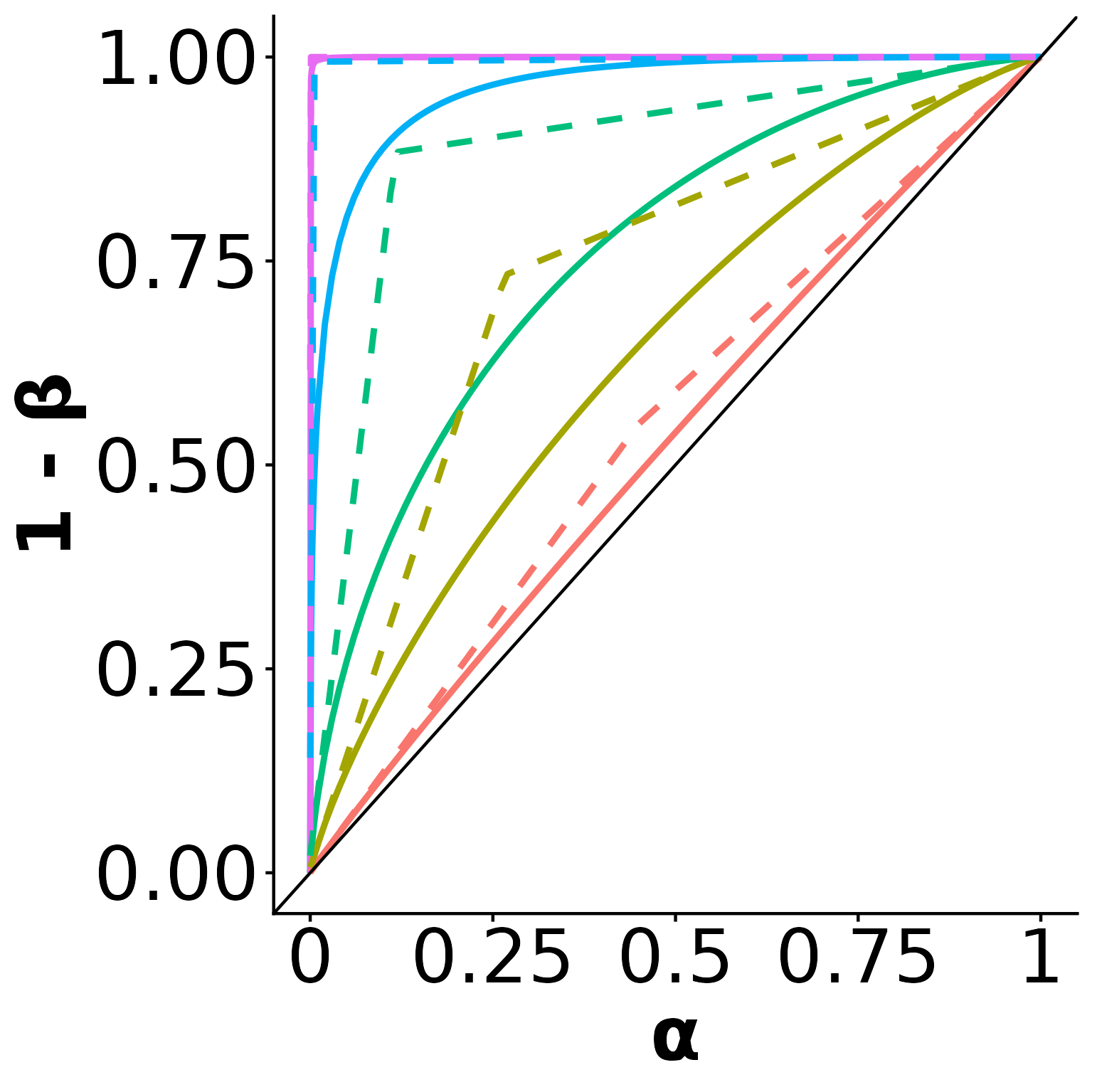} &
     \includegraphics[width=.33\linewidth]{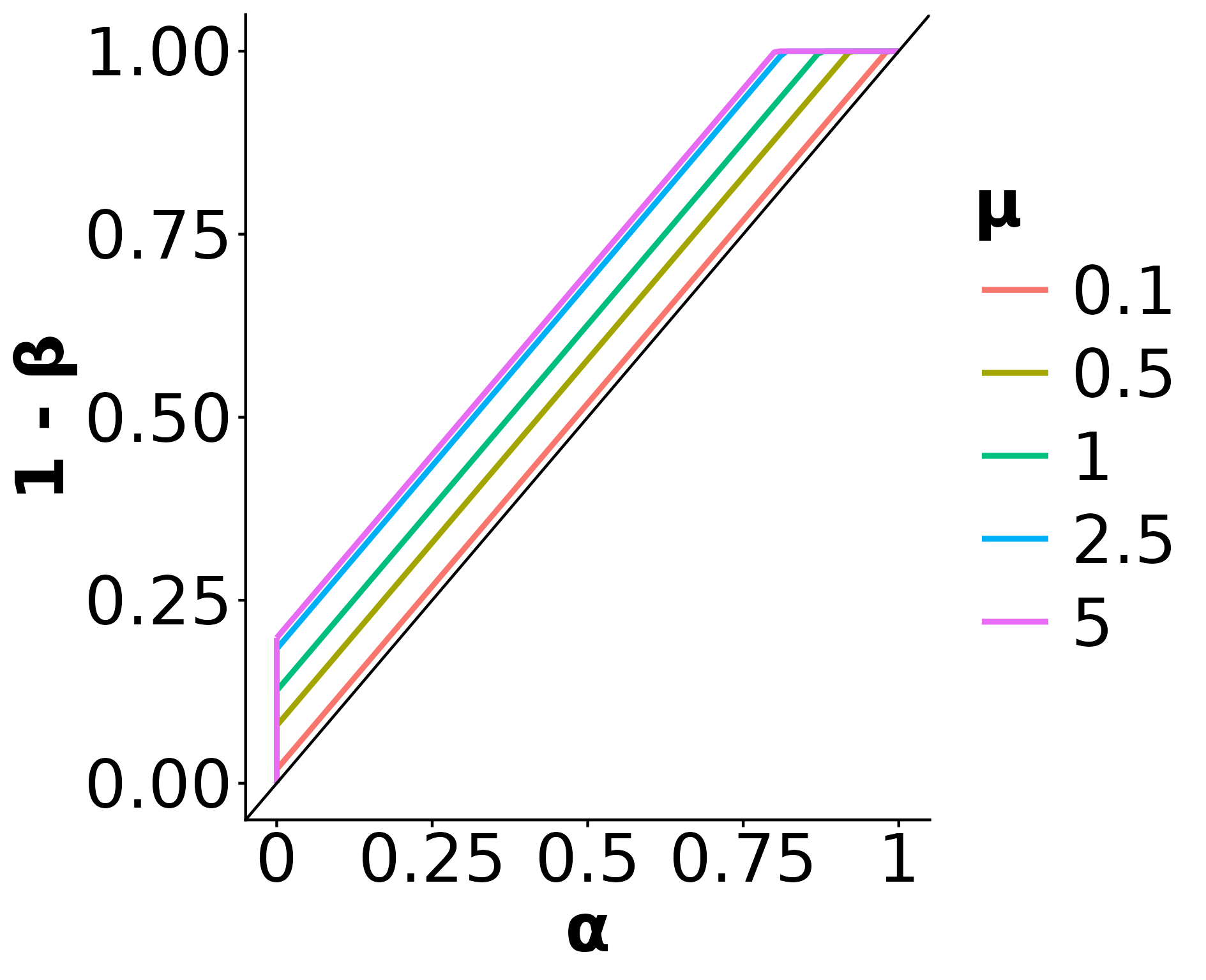}
\end{tabular}
\caption{Trade-off functions (solid) and pure, resp. approx, DP bounds (dashed) for three classical $f$-DP mechanism: Laplace (left), Gaussian (center) and Uniform Random Sampling (right) with $\sup_{D,D'} |q(D) - q(D')| = 1$ and $n= 5$.}
\label{fig: roc-curves}
\end{figure}

As illustrated in Figure \ref{fig: roc-curves}, ROC curves that lie closer to the upper-left corner indicate a greater ability to minimize both types of errors simultaneously. Therefore, bounding the distinguishability between $H_0$ and $H_1$ amounts to placing an upper bound on the ROC curves of all tests attempting to differentiate between hypotheses in \eqref{eq: hyp f-dp}. This is precisely the guarantee provided by the $f$-Differential Privacy framework \citep{Dong2022}.

\begin{definition}
    A mechanism $Q$ with privacy-loss parameters loss $\mu$ is $f_\mu$-DP if for all neighboring datasets $D$ and $D'$ and all possible test statistics $T$, there exists a function $f: [0,1] \rightarrow [0,1]$ such that all the ROC-curves of the test statistics $T$ satisfy
    $$
    1 - f_\mu(\alpha) \geq 1 - \beta_{T}(\alpha).
    $$
    The function $f_\mu$ is called a trade-off function and satisfies $f_\mu(\alpha) = \inf_{T,\phi}\{\beta_T(\alpha_\phi) \leq \alpha\}$.
\end{definition}
This definition formalizes privacy in terms of the existence of an upper bound on adversarial inference performance, captured by the trade-off function $f_\mu$. Its existence is guaranteed by the Neyman–Pearson lemma \cite[p.66]{Lehmann2022}, which also provides a constructive method for identifying the tightest possible bound. Specifically, the most powerful test, the one achieving the trade-off bound, is given by the likelihood ratio test \cite[p.17]{Lehmann2022}.
For many standard $f$-DP mechanisms, such as the ones below, the trade-off function can thus be explicitly computed.

\begin{example}\label{ex: lap}
    Given a dataset $D$, a query $q$, and a privacy-loss parameter $\mu > 0$, the {\bf Laplace mechanism} adds noise as follows: 
    $$
    Q(D) = q(D) + Lap\left\{\frac{\sup_{D,D'} |q(D) - q(D')| }{\mu}\right\},
    $$
    where $Lap(\sigma)$ denotes a Laplace random variable with zero mean $0$ and scale  $\sigma$. %Here, $\Delta q = \sup_{D,D'} |q(D) - q(D')| $ is the sensitivity of the query $q$.
\end{example}

\begin{example}\label{ex: gaus}
    For a dataset $D$ and a query $q$, the {\bf Gaussian mechanism} is defined by
    $$
    Q(D) = q(D) + N\left\{0, \frac{\sup_{D,D'} |q(D) - q(D')| }{\mu}\right\},
    $$
    where $N(\sigma)$ denotes a Gaussian random variable with zero mean $0$ and scale  $\sigma$.%Again, $\Delta q = \sup_{D,D'} |q(D) - q(D')| $ is the sensitivity of the query $q$.
\end{example}

\begin{example}\label{ex: unif}
    For a dataset $D$ of size $n > 0$, the {\bf Uniform Random Sampling mechanism} with privacy loss parameters $\mu > 0$ is defined as
    $$
    \begin{array}{ll}
        c, & \text{with probability } e^{-\mu},
        \\
        x_i \in D, & \text{with probability } \frac{1 - e^{-\mu}}{n},
    \end{array}$$
    where $x_i$ is selected uniformly at random within $D$ and $c \in \mathbb{R}$ is a fixed constant, independent of $D$, distinguishable from any possible record value $x_i \in D$.
    %and the privacy-loss parameter is $\mu = p / n$, with $n >0$ denoting the number of records in $D$.
\end{example}

Figure \ref{fig: roc-curves} presents the ROC curves for Examples \ref{ex: lap}, \ref{ex: gaus}, and \ref{ex: unif} under various values of privace-loss parameter $\mu$.
For noise-addition mechanisms based on log-concave distributions, such as Gaussian and Laplace, the trade-off function is
$$
f_\mu(\alpha) = F\left\{F^{-1}(\alpha) - \mu\right\},
$$
where $F$ is the cumulative distribution function (CDF) of a zero-mean, unit-scale continuous random variable taking values in $\mathbb{R}$ \citep{Dong2022}.
%The derivation of the trade-off function for Example \ref{ex: unif} is provided in the Appendix.

The $f$-DP framework encompasses classical definitions of privacy, including pure differential privacy in \eqref{eq: pure dp} and its relaxation, known as $(\epsilon,\delta)$-DP or approximate DP. The latter guarantees that \eqref{eq: pure dp} holds with probability at least $1 - \delta$. However, approximate DP offers no specific guarantees about the mechanism’s behavior on the residual $\delta$-probability mass.
Any mechanism satisfying pure or approximate DP has its ROC-curve function bounded above by \begin{equation}\label{eq: tf pure dp}
    1 - \max\{0, 1- \delta - e^\epsilon \alpha, e^{-\epsilon} (1 - \delta - \alpha)\},
\end{equation}
with the pure DP case covered by setting $\delta = 0$ \citep{Dong2022}.

Shown in Figure \ref{fig: roc-curves} by dashed lines, these bounds are generally looser than those provided by $f$-DP.
%, particularly in the context of hypothesis testing.
Nonetheless, they remain valuable for summarizing privacy guarantees in terms of a small set of interpretable numerical parameters.
Interestingly, the Gaussian mechanism, which does not satisfy the strict inequality of \eqref{eq: pure dp}, is often viewed as failing to offer strong privacy guarantees. Within the $f$-DP framework, however, hypothesis testing offers a richer interpretation of this failure allowing us to understand precisely its protection guarantees; see Section \ref{sec: unbounded}.

Figure \ref{fig: roc-curves} also highlights a limitation of ROC curves in privacy evaluation. For instance, the Uniform Random Sampling mechanism may yield ROC curves that remain close to the diagonal, traditionally interpreted as low distinguishability, while in reality being prone to \emph{catastrophic} failures through blatant disclosure of individual data points.
This demonstrates that ROC curves, and their widely-used summary statistic, the Area Under the Curve (AUC), can be misleading indicators of privacy. Although AUC, defined as the integral of the ROC curve over the interval $[0;1]$, is a standard metric in statistics and machine learning, it fails to capture key aspects of disclosure risk in the privacy context.
Consequently, ensuring strong privacy guarantees require additional conditions on the ROC curve as well as alternative measures of disclosure risk. We address this issue through a Bayesian framing of hypothesis testing.

\subsection{Bayesian Statistics and Prior-to-Posterior Analysis}\label{sec: prior-to-posterior}
Bayesian statistics integrates prior beliefs or knowledge into analysis by using Bayes’ theorem to update prior probabilities into posterior distributions. This approach is particularly valuable because it allows for a principled quantification of the useful information contained in the data by comparing the prior and posterior distributions.
In what follows, we focus on the application of Bayesian statistics to hypothesis testing. For a broader treatment of Bayesian methods across other domains of statistics, we refer the reader for instance to \cite{Gelman2020}.

\begin{definition}
    A {\bf prior probability} $p_{\text{prior}} \in [0,1]$ is the probability
    $$
    p_{\text{prior}}= \text{Pr}\{H_1\}.
    $$
    that hypothesis $H_1$ holds true \emph{before} performing any statistical test.%, i.e., before computing the test statistic $T$ and comparison with the threshold $\phi$.
\end{definition}
%In practice, the prior probability $p_{\text{prior}}$ may take any value in the interval $[0;1]$. To assess the distinguishability between $H_0$ and $H_1$ in full generality, it is necessary to consider the entire range of possible prior probabilities. Although one may choose to restrict this range, doing so is not recommended: such a restriction would effectively impose assumptions about the availability or credibility of auxiliary information, which are often highly contextual and may evolve over time.
Common examples of prior probabilities include: 
\begin{itemize}
    \item Random guess ($p_{\text{prior}} = 0.5$): This prior represents a state of complete uncertainty, in which no auxiliary information is available before the test.% and performance is no better than random guessing between the two hypotheses.
    \item Prevalence:  Borrowed from epidemiology, prevalence denotes the proportion of a population exhibiting a particular characteristic or condition. It is often publicly available at national or regional levels and can inform prior probabilities when the attacker knows the geographic location of an individual.
\end{itemize}
%More generally, any auxiliary information about the target privacy unit, or group, available before the attack can be used to obtain a prior probability.

Hence, the prior captures the baseline level of information available for distinguishing between $H_0$ and $H_1$. Once the disseminated query results $q$ is used, typically via a test statistic $T$, the prior is updated to a posterior probability.
\begin{definition}
    The {\bf posterior probability} $p_{\text{post}} \in [0,1]$ is the probability 
    $$
    p_{\text{post}}= \text{Pr}\{H_1 | T > \phi\}.
    $$
    that hypothesis $H_1$ holds true \emph{after} a positive test result.%, i.e., after computing the test statistics $T$ and observing that it exceeds the threshold $\phi$.
\end{definition}
The posterior probability of a successful test can be computed via Bayes’ theorem: 
\begin{equation}\label{eq: bayes}
p_{\text{post}} = \frac{\text{Pr}\{T > \phi | H_1\}}{\text{Pr}\{T > \phi \}} \times p_{\text{prior}}.
\end{equation}
Equation \eqref{eq: bayes} provides a means to quantify the amount of useful information contained within $q$ to help distinguish between the two hypothesis when $T > \phi$. In the extreme case where $p_{\text{post}} = 1$, the hypothesis $H_0$ is rejected with certainty, resulting in perfect distinguishability between datasets $D$ and $D'$. This scenario corresponds to a blatant disclosure of private information.
Controlling the extent to which the posterior probability can approach $1$, and thereby regulating the inferential power gained through the test, is critical to maintaining privacy. These issues are addressed in detail in Section \ref{sec: privacy garantee}.

%}

%\showmatmethods{} % Display the Materials and Methods section

\section{Managing disclosure risk under uncertainty}\label{sec: privacy garantee}

\subsection{Membership Attacks}
%We begin by framing membership attacks as statistical hypothesis tests targeting a dataset $D$ composed of $n$ records $\{x_i\}_{i = 1,\dots,n}$.
We begin by framing membership attacks as statistical hypothesis tests targeting a dataset $D$ composed of $n$ records $\{x_i\}_{i = 1,\dots,n}$, under the Bayesian prior-to-posterior formulation introduced in Section \ref{sec: prior-to-posterior}. In this context, it is necessary to distinguish between two kinds of prior updates.
The first kind does not depend on an individual's presence in the dataset. Consider an attacker uncertain whether Alice, a long-term smoker, has lung cancer. Suppose an independent study, conducted on an unrelated dataset, establishes that smoking causes lung cancer. The attacker's belief regarding Alice's condition would rise substantially, irrespective of whether her data ever entered any dataset \citep[Ch. 8 §1.1.2]{kifer2011nofreelunch, kasiviswanathan2014semantics, drechsler2021handbook}. This form of belief revision, sometimes termed inferential or generalization risk, is a property of the statistical relationship being disclosed rather than of the specific data used to estimate it. The relevant question is not what the released numbers reveal about Alice individually, but what they reveal about individuals sharing her characteristics, and whether disclosing them is warranted given their nature, irrespective of whose records happened to be collected to produce them.
The second kind of prior update depends specifically on an individual's own presence in the collected data. This is the notion of privacy under study throughout this paper: we consider only the change in belief attributable to the presence, within the collected dataset, of the target individual's own record.
This restriction is deliberate. Data protection regulations, such as the GDPR or the Swiss LPD, impose obligations on the processing of collected personal data, rather than on the dissemination of general statistical relationships. Correspondingly, this narrower prior-to-posterior view is the one a data protection authority is positioned to act upon: when approving the creation and dissemination of a query output under a given privacy-loss budget, the authority assesses the risk incurred by the individuals whose data was collected to produce it, not the broader and largely uncontrollable risk of statistical generalization.

For simplicity, we assume a one-to-one correspondence between records and privacy units; that is, each individual contributes exactly one record to the dataset. This assumption is made without loss of generality, as all subsequent results can be extended to scenarios where individuals may contribute a bounded number of records; see \cite[p. 230]{Dwork2014}.

We further assume that $D$ exclusively consists of records belonging to a sensitive group of individuals; thus, inclusion in $D$ directly implies group membership. In practice, datasets often span multiple groups, such as when representing an entire population. Membership attacks can easily target specific subgroups defined by any categorical variable that partitions the dataset, e.g., gender, diagnosis, region.
%defines a group whose membership status can be attacked.
We also assume that $m >0$ queries, denoted $\{q_j(D)\}_{j = 1,\dots,m}$, are available for the attacker. For simplicity, we refer to the vector of disseminated queries as $q$, dropping the index notation.

\begin{definition}
Given a dataset $D$ and disseminated queries $q$, a {\bf membership attack} is an algorithm that attempts to determine whether a specific record $x_{target}$ was included in the dataset $D$ used to generate $q$. 
\end{definition}

Our objective is to evaluate the effectiveness of membership attacks under assumptions on the attacker’s knowledge and capabilities that are chosen to ensure the robustness of the derived guarantees. To this end, we adopt a conservative, worst-case scenario in which the attacker knows the status and data of all individuals in $D$ except for one. Although this assumption may appear strong, it is essential to establish a rigorous connection between privacy guarantees and the success of membership attacks. Any relaxation of this assumption would necessitate additional hypotheses about the nature and extent of auxiliary information available to the attacker, which would in turn undermine the robustness and generality of the privacy guarantee.
Such auxiliary information might indeed be widely available post hoc, rendering the data release retrospectively unsafe and voiding the "future-proof" assurances provided by differential privacy.

Despite its strength, this assumption is not unrealistic. In many real-world scenarios, especially in health or demographic studies, external actors may indeed possess near-complete knowledge of the population, with only a few uncertain records. Examples include: HIV-positive status known to a major health insurance provider,
diabetes status inferred from grocery shopping patterns,
cardiovascular conditions derived from wearable device data,
financial status inferred from online banking activity,
household composition extracted from census data.
Although this assumption is required for the theoretical analysis, it is worth emphasizing that membership attacks in practice can succeed under significantly weaker knowledge assumptions \citep{Dwork2017}.

The principle underlying membership attacks is intuitive: consider the dataset
$$
D_{-n} = \{x_1, \dots, x_{n -1}\},
$$
and compare the disseminated output $q$ with the output that would result from adding the target record, i.e., $q(D_{-n} \cup \{x_{target}\})$.
If $q$ is disseminated exactly, i.e., without randomization, and if $q = q(D_{-n} \cup \{x_{target}\}) $, then it becomes trivial to infer that $x_{target} \in D$. Under such conditions, the membership attack succeeds with certainty.
The only viable strategy to mitigate this vulnerability is to release $q$ through a randomized mechanism, i.e., to publish $Q(D)$ instead of the exact query result.

Assuming $q$ is the realization of a randomized mechanism $Q$, a direct comparison between $q$ and $Q(D_{-n} \cup \{x_{target}\})$ becomes statistically uncertain.
Consequently, the attacker frames the problem as a hypothesis test: 
\begin{equation}
    \begin{array}{ll}
         H_0: & Q \stackrel{d}{=} Q(D_{-n}),  \\
         H_1: & Q \stackrel{d}{=} Q(D_{-n} \cup \{x_{target}\}),
    \end{array}
\end{equation}
where $\stackrel{d}{=}$ denotes equality in distribution.
Using an appropriate test statistic, the attacker attempts to distinguish between $H_0$ and $H_1$ with a given level of confidence.

In this framework, the standard notions of hypothesis testing acquire new interpretations:
\begin{itemize}
    \item A Type I error $\alpha$ corresponds to a false positive, where the attacker incorrectly infers that $x_{target}\in D$ when in fact it is not.
    \item A Type II error $\beta$ corresponds to a false negative, where the attacker fails to detect the inclusion of $x_{target}\in D$ despite its presence in $D$.
\end{itemize}
    
The selection of the threshold $\phi$, and consequently the values of $\alpha$ and $\beta$, is part of the attacker’s strategy. Naturally, the attacker aims to minimize both $\alpha$ and $\beta$; however, as established in Section \ref{sec: f-dp}, the $f$-DP framework constrains this optimization via the trade-off function that imposes a fundamental balance between the two error rates.
Two illustrative strategies are:
\begin{itemize}
    \item Setting $\alpha = 0$ (or as close as possible) ensures that any positive test result corresponds to a true member. This is ideal in targeted attacks involving a large pool of candidates: although many true members may be missed, successful detections are definitive and actionable.
    \item Setting $\beta = 0$ (or as close as possible) guarantees detection of all true members, but may include numerous false positives. This strategy is relevant in scenarios where missing a true member is more costly than falsely identifying non-members, for example, in loan approval processes where avoiding default is prioritized over excluding a few creditworthy applicants.
\end{itemize}
In either case, the attacker’s ultimate objective is to increase their confidence that $x_{target}\in D$, i.e., to make their posterior belief in $H_1$ as close  as possible to $0$ or $1$. This raises the critical question: To what extent can the disseminated query results strengthen the attacker’s belief in the target's membership?

To address this, we leverage the Bayesian formulation of hypothesis testing and integrate it with the $f$-differential privacy framework. This allows us to formally characterize, and crucially control via privacy-loss parameters, the degree to which a membership attack can reinforce an attacker’s belief in the inclusion of a target individual.

\subsection{Controlling membership attacks with $f_\mu$-DP}\label{sec: privacy}
Assume that the attacker holds a prior belief $p_{prior}$ regarding the plausibility that the target individual is a member of the sensitive group under investigation.
This prior probability $p_{\text{prior}}$ may take any value in $(0;1)$. To assess the distinguishability between $H_0$ and $H_1$ in full generality, it is necessary to consider the entire range of possible prior probabilities. Although one may choose to restrict this range, doing so is not recommended: such a restriction would effectively impose assumptions about the availability or credibility of auxiliary information, which are often highly contextual and may evolve over time.

The disclosure risk of a membership attack using the DP-protected responses $q$ can be evaluated through Bayesian hypothesis testing.
Beyond the nature of the DP-mechanism and its parameters, disclosure risk also depends the attacker’s strategy, particularly the choice of test threshold $\phi$, which lie outside the data curator’s control.
Hence, a mechanism with strong privacy guarantees provides a high level of protection regardless of the prior probability $p_{\text{prior}}$ or the choice of threshold $\phi$, equivalently of the confidence level $\alpha$.

\subsubsection{Mechanisms suffering from catastrophic failures}\label{sec: unbounded}
A membership attack discloses the target's membership status with certainty if the attacker can elevate the posterior belief to $p_{posterior} = 1$.
We refer to such a disclosure as blatant. If a randomized mechanism fails to prevent blatant disclosures, it is deemed vulnerable and does not provide strong privacy guarantees.

\begin{definition}\label{def: failures} 
 A randomized mechanism $Q$ is said to fail if for all $p_{prior} \in [0;1]$
 \begin{equation*}
     \sup_{\alpha} p_{\text{posterior}} 
     = %\sup_{\alpha} \frac{p_{\text{prior}}\{1 - f_\mu(\alpha)\}}{(1 - p_{\text{prior}}) \alpha + p_{\text{prior}} \{1 - f_\mu(\alpha)\}} =
     1.
 \end{equation*}
 Moreover, if for all $\alpha$ and $p_{\text{prior}}\in [0;1]$
    $$
    p_{\text{posterior}} < 1,
    $$
    then the mechanism $Q$ is said to \emph{fail gracefully}. Otherwise the mechanism $Q$ is said to \emph{fail catastrophically}.
\end{definition}
The posterior probability $p_{\text{posterior}}$ can equivalently be expressed using Bayes' theorem (4), by substituting the trade-off function  $f_\mu$ for 
$\text{Pr}\{T > \phi | H_1\}$, and applying the law of total probability on $\text{Pr}\{T > \phi \}$, which yields:
$$
p_{\text{posterior}} 
     =  \frac{p_{\text{prior}}\{1 - f_\mu(\alpha)\}}{(1 - p_{\text{prior}}) \alpha + p_{\text{prior}} \{1 - f_\mu(\alpha)\}}.
$$

Mechanisms prone to catastrophic failure are vulnerable to blatant disclosures, as they admit a successful membership attack that can reveal a sensitive attribute with certainty. A canonical example is the Uniform Random Sampling Mechanism discussed in Example \ref{ex: unif} where with probability $(1 - e^{-\mu})$, a dataset record is disclosed. While this does not result in a systematic privacy breach, the likelihood of such an event is far from negligible.
In contrast, graceful failures allow the attacker to raise their confidence arbitrarily close to $1$, but a non-zero residual uncertainty always remains.
Theorem \ref{th: failure conditions} provides a formal condition under which a mechanism fails catastrophically, i.e., it admits distinguishing events, that allow the attacker to determine the target's membership status with certainty.
\begin{theorem}\label{th: failure conditions}
    A randomized mechanism $Q$ with trade-off function $f_\mu$ is failing catastrophically if and only if 
    $$
    f_\mu(0) < 1.
    $$
    Alternatively, a gracefully failing mechanisms satisfy for all $\alpha > 0$
    $$
    f_\mu(\alpha) < 1 \text{ and } \lim_{\alpha\rightarrow 0} f_\mu(\alpha) = 1.
    $$
\end{theorem}

%Theorem \ref{th: failure conditions} provides a formal condition under which a mechanism admits distinguishing events—outcomes that allow the attacker to determine the target's membership status with certainty. This occurs, for instance, in Example \ref{ex: unif}, where with probability $p$, a dataset record is directly disclosed. While this does not result in a systematic privacy breach, the likelihood of such an event is far from negligible.
%It is important to note that Theorem \ref{th: failure conditions} offers a necessary but not sufficient condition on the trade-off function for graceful failure. A complete characterization will be presented in Section \ref{sec: bounded}.

\begin{figure}%[tbhp]
\centering
\begin{tabular}{c}
     \includegraphics[width=0.95\linewidth]{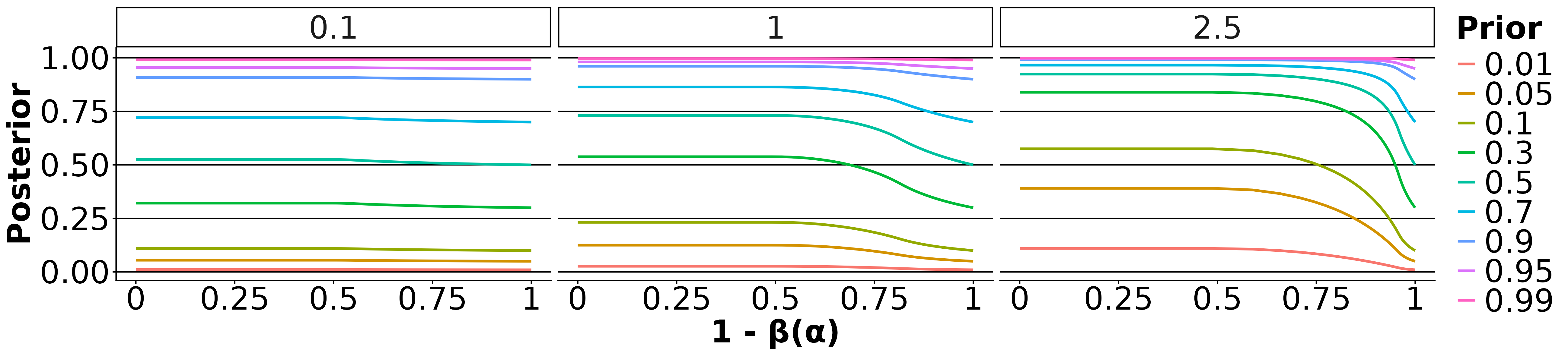} \\
     \includegraphics[width=0.95\linewidth]{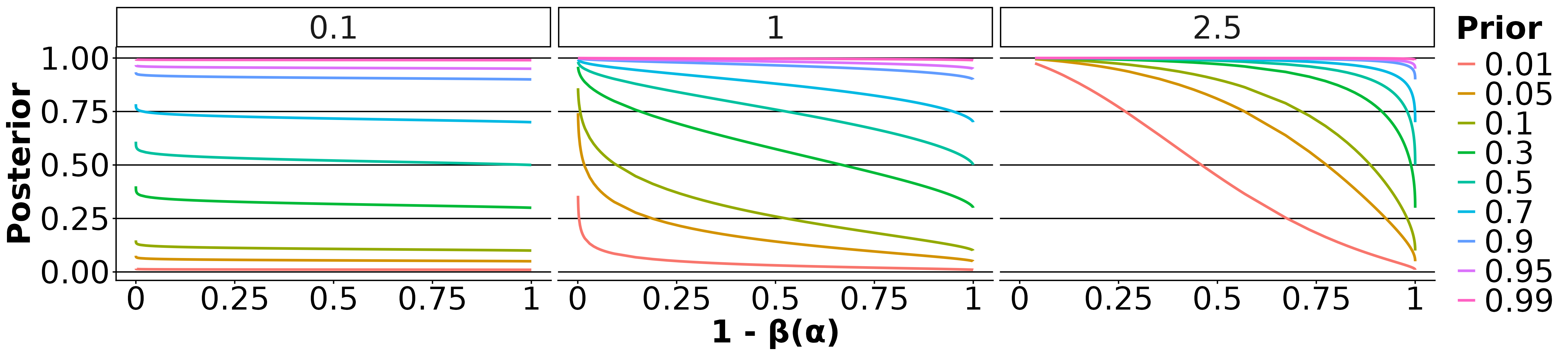} \\
     \includegraphics[width=0.95\linewidth]{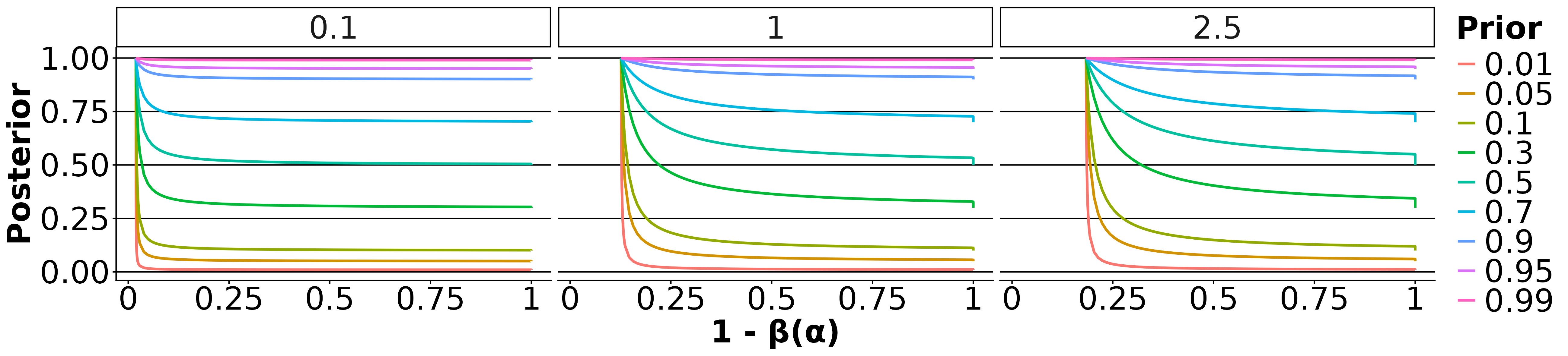}
\end{tabular}
\caption{Posterior probability obtained by updating the prior probability following a successful membership attack on queries protected using Laplace (top row), Gaussian (middle row) and Uniform Random sampling (bottom row) mechanisms for privacy-loss parameters $\mu = 0.1$ (left), $1$ (middle), $2.5$ (right). %The posterior probability is shown here as function of $1 - \beta(\alpha)$ for explicit visualization of blatant disclosure.
}
\label{fig: posterior prob}
\end{figure}

Figure \ref{fig: posterior prob} illustrates posterior probabilities obtainable using the best performing test statistic for the three mechanisms introduced in Section \ref{sec: f-dp}:
\begin{itemize}
    \item The Uniform Random Sampling mechanism fails catastrophically, as its trade-off function meets the criteria of Theorem \ref{th: failure conditions}. %Indeed, the figure confirms that for certain $\alpha$ values, the posterior probability reaches exactly $1$.
    \item The Gaussian mechanism represents a graceful failure: posterior probabilities can approach 1 but never reach it, maintaining a residual uncertainty independently of $p_{prior}$.
    \item The Laplace mechanism does not fail as its posterior probabilities remain strictly below $1$ and is bounded from above for each $p_{prior}$.
\end{itemize}
In practice, when data is disseminated to untrusted data recipients, we strongly recommend avoiding mechanisms that fail catastrophically, as they offer insufficient privacy guarantees.
%Beyond the choice of mechanism, vulnerability also depends the attacker’s parameter choices—particularly the test threshold $\phi$—which lie outside the data curator’s control. For this reason, a mechanism with strong privacy guarantee offers good level of protection independently of the choice of $\alpha$.
Catastrophic failure is not merely a theoretical concern: it arises whenever a mechanism releases some outputs exactly, even at random. A concrete example is a variant of the exponential mechanism for private quantile release that selects its output at random from among the values already present in the dataset, with probability increasing as the candidate approaches the true empirical quantile. Because only values already present in $D$ can be selected, observing an output equal to the target individual's value discloses their membership with certainty, an event that occurs with strictly positive probability, and disproportionately so for individuals whose value lies close to the quantile of interest. This vulnerability is a consequence of restricting the candidate outputs to the dataset itself, rather than a necessary feature of the exponential mechanism: standard formulations avoid it either by selecting from a predefined candidate grid, independent of $D$, \citep{mcsherry2007mechanism} or by returning a value drawn uniformly at random between two dataset points \citep{smith2011privacy}, so that no output can be attributed with certainty to any single record.

Figure \ref{fig: posterior prob} shows a precision–recall curve, a standard tool for evaluating binary classifiers under class imbalance: precision, the probability that an individual flagged as a member by the attack is a true member, corresponds to the posterior probability $p_{\text{posterior}}$, while recall, the proportion of true members correctly detected by the attack, corresponds to its power $1 - f_\mu(\alpha)$. As with any classifier, the closer the curve lies to the top-right corner, the stronger the attack: high precision and recall simultaneously indicate that the attacker can detect most members while rarely misclassifying a non-member, whereas a curve that remains close to the prior baseline across recall values reflects an attack that gains little from the observed output. In Figure \ref{fig: posterior prob}, the posterior probability coincides with the prior at $\beta = 0$, offering a helpful visual benchmark: As $\beta$ increases, the gap between prior and posterior probabilities widens, reflecting the amount of information the attacker gains from observing $q$ and applying a corresponding test statistic resulting in a successful attack. This behavior enables a refined analysis of the privacy guarantees offered by mechanisms such the Gaussian and Laplace mechanisms.

%Finally, we need to quantify the guarantee offered by mechanism that gracefully fails, i.e., with unbounded relative disclosure risk but strictly less than $1$ posterior probability.
%In case of risk boundeness, we have seen that the relative disclosure risk is bounded by $e^\epsilon$.

%Thanks to trade-off functions, we can see that such condition holds in probability and we can find the probability for which the relative disclosure risk remains below such value; see black dashed line on Figure \ref{}, b).
%The probability for which the risk is bounded is given by
%\begin{equation}\label{eq: prop bounded risk}
%\pi = \inf_{\alpha} \left\{\alpha: \frac{1 - f(\alpha)}{\alpha} < e^\epsilon\right\}.    
%\end{equation}
%Equation \eqref{eq: prop bounded risk} might not be solvable in closed for but numerical solver can provide robust estimate of the lower bound.
%Table \ref{tab: } gives the value of $\pi$ for the Gaussian mechanism and multiple choices of $\epsilon$: we observe that ...

%We now three interpretable measures to quantify disclosure risk, characterize randomized mechanisms and link them directly to the budget through the notion of relative disclosure risk.

\subsubsection{Interpreting disclosure risk}\label{sec: bounded}
We now turn our attention to randomized mechanisms that do not fail catastrophically. In such cases, we analyze the behavior of the ratio between the posterior and prior probabilities to quantify the amount of information that the disseminated query results $q$ convey to an attacker in the context of a membership inference attack.

\begin{definition}\label{def: rel risk}
    The maximum relative disclosure risk $\rho$ for the mechanism $Q$ with trade-off function $f_\mu$ is given by
    $$
    \rho = \sup_{\alpha, p_{\text{prior}}} \frac{p_{\text{posterior}}}{p_{\text{prior}}} = \sup_{\alpha \in [0;1]} \frac{1 - f_\mu(\alpha)}{ \alpha}.
    $$
\end{definition}
The notion of relative disclosure risk originates in epidemiology, where it compares the probability of an event, e.g., contracting a disease, occurring in an exposed group to the same probability in an unexposed group. In our setting, it quantifies the likelihood of the target's membership increases due to the release of query results $q$.
Relative disclosure risk thus serves as a privacy-loss metric, analogous to the privacy loss budget $\epsilon$, and provides an interpretable measure of the potential gain in an attacker's confidence.
Specifically, it quantifies the maximum increase in belief that an adversary can attain through a membership attack using $q$.

This leads to a classification of mechanisms into two broad categories: those that offer bounded relative disclosure risk, i.e., they strictly limit the adversary’s gain in confidence, and those that offer only guarantees for a limited range of type I errors values $\alpha$.

\begin{theorem}\label{th: bounded risk}
A mechanism $Q$ with continuous trade-off function $f_\mu$ has bounded relative disclosure risk $\rho > 0$ if and only if 
\begin{equation}\label{eq: th1}
    \lim_{\alpha \rightarrow 0} \frac{1 - f_\mu(\alpha)}{\alpha} < \infty.
\end{equation}
In particular, \eqref{eq: th1} holds true when there exists $\rho \geq 1$ such that $f_\mu(\alpha) \sim 1 - \rho\times \alpha$, where $\sim$ is the asymptotic equivalence when $\alpha \rightarrow 0$. 
\end{theorem}

\begin{corollary}\label{cor: tail df}
    For any noise addition mechanism $Q$ where the added noise $N$ is a log-concave random variable taking value in $\mathbb{R}$ with cumulative distribution $F$, the relative disclosure risk is bounded if, and only if, the tail distribution function $\bar{F} = 1 - F$ satisfies
    $$
     \lim_{\alpha \rightarrow 0} \frac{\bar{F}(F^{-1}(1 - \alpha) - \mu)}{\bar{F}(F^{-1}(1 - \alpha))} < \infty.
    $$
\end{corollary}

Together, Theorem \ref{th: bounded risk} and Corollary \ref{cor: tail df} provide sufficient conditions under the $f$-DP framework for a mechanism to offer strong, interpretable privacy guarantees.
Corollary \ref{cor: tail df} is a simple restatement of Theorem \ref{th: bounded risk}, yet it remains useful, as it highlights the crucial role of tail behavior: if the noise distribution decays too rapidly in the tails, the mechanism cannot ensure bounded relative disclosure risk.

Interestingly, these conditions align naturally with the guarantees offered by pure $\epsilon$-differential privacy.
\begin{corollary}\label{cor: pure dp}
    Any randomized mechanism $Q$ that satisfies pure $\epsilon$-DP also satisfies the conditions for Theorem \ref{th: bounded risk} with 
    $$
    \rho \leq e^\epsilon.
    $$
\end{corollary}

\begin{figure}%[tbhp]
\centering
\begin{tabular}{c}
     \includegraphics[width=0.95\linewidth]{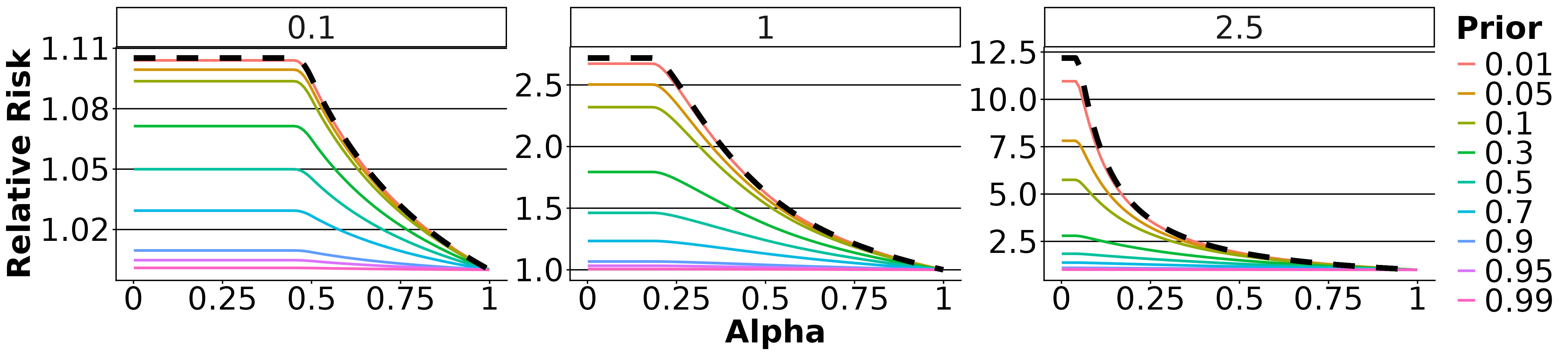} \\
     \includegraphics[width=0.95\linewidth]{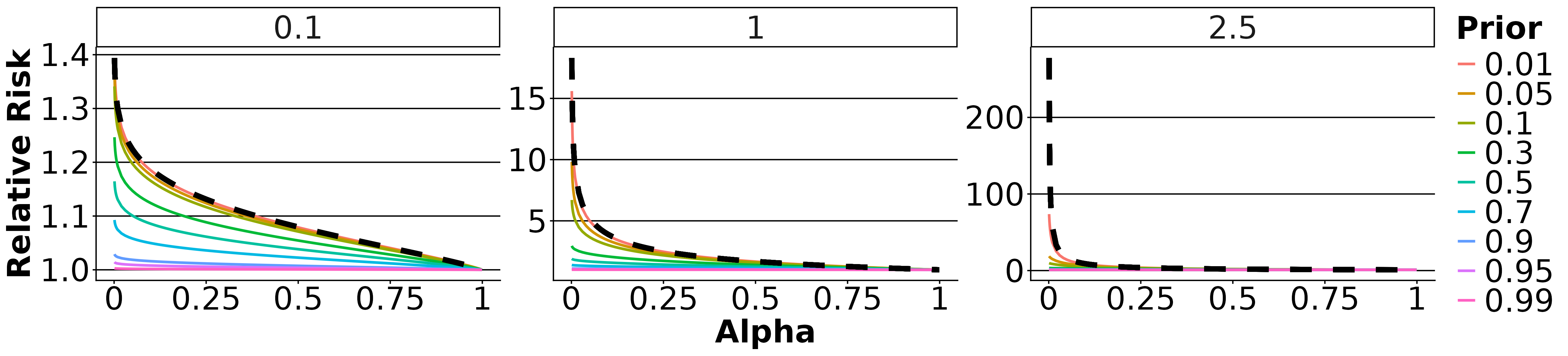}
\end{tabular}
\caption{Relative disclosure risk for the Laplace (top) and Gaussian (bottom) mechanisms as function of $\alpha$ for privacy-loss parameters $\mu = 0.1$ (left), $1$ (middle), $2.5$ (right).}
\label{fig: relative disclosure risk}
\end{figure}

This result establishes a direct correspondence between mechanisms that satisfy pure $\epsilon$-DP and those with bounded relative disclosure risk under $f$-DP.
The Laplace mechanism, discussed in Example \ref{ex: lap}, meets these criteria and satisfies $\epsilon = \mu =\log \rho$, establishing a direct link between all these parameters; see Lemma A.6 in the appendices of \cite{Dong2022} for explicit formulas of the trade-off function, which includes a linear regime for sufficiently small $\alpha$.
In this setting, $\rho$ provides a privacy metric quantifying the largest possible decrease in uncertainty brought by a membership attack, with a direct correspondence to the privacy-loss budget $\epsilon$.
Figure \ref{fig: relative disclosure risk} presents the relative disclosure risk for both the Laplace and Gaussian mechanisms under various privacy-loss parameters. As expected, the Laplace mechanism yields bounded relative disclosure risk.
%In contrast, Gaussian noise exhibits qualitatively different behavior.

In contrast, the Gaussian mechanism, like the Uniform Random Sampling, has unbounded relative disclosure risk, i.e., $\rho = \infty$.
Nevertheless, the former still provides meaningful privacy protection.
As discussed in Section \ref{sec: unbounded}, the Gaussian mechanism does not allow for blatant disclosures as it fails only gracefully since an attacker’s confidence never reaches exactly $1$.
Blatant disclosure would require to set $\alpha=0$, corresponding to an infinite threshold $\phi$ on the test statistic, which is completely uninformative.
Thus, it is meaningful to restrict the analysis to realistic thresholds values corresponding to analyse disclosure risk for $\alpha > \alpha_0 > 0$.
%In practice, standard choices for $\alpha$ in hypothesis testing are $0.05$ or $0.01$, targeting rejection rates of $95\%$ or $99\%$, with desired test power $1 - \beta$ exceeding $0.8$.

Accordingly, we define the relative disclosure risk bound for confidence level at most $\alpha_0$ as:
$$
\rho_{\alpha_0} =  \sup_{\alpha \in [\alpha_0;1]} \frac{1 - f_\mu(\alpha)}{\alpha}.
$$
In practice, standard choices for $\alpha$ in hypothesis testing are $0.05$ or $0.01$, targeting rejection rates of $95\%$ or $99\%$ and a test power $1 - \beta$ exceeding $0.8$.
A conservative analysis might thus set $\alpha_0 =10^{-2}$ or even $10^{-3}$. For noise-addition mechanisms with log-concave distributions, this yields
$$
\rho_{\alpha_0} = \frac{1 - {F}\{F^{-1}(1 - \alpha_0) - \mu\}}{\alpha_0},
$$
which provides a direct link between the privacy-loss parameter $\mu$ and the relative disclosure risk $\rho_{\alpha_0}$.

\begin{figure}%[tbhp]
\centering
\includegraphics[width=\linewidth]{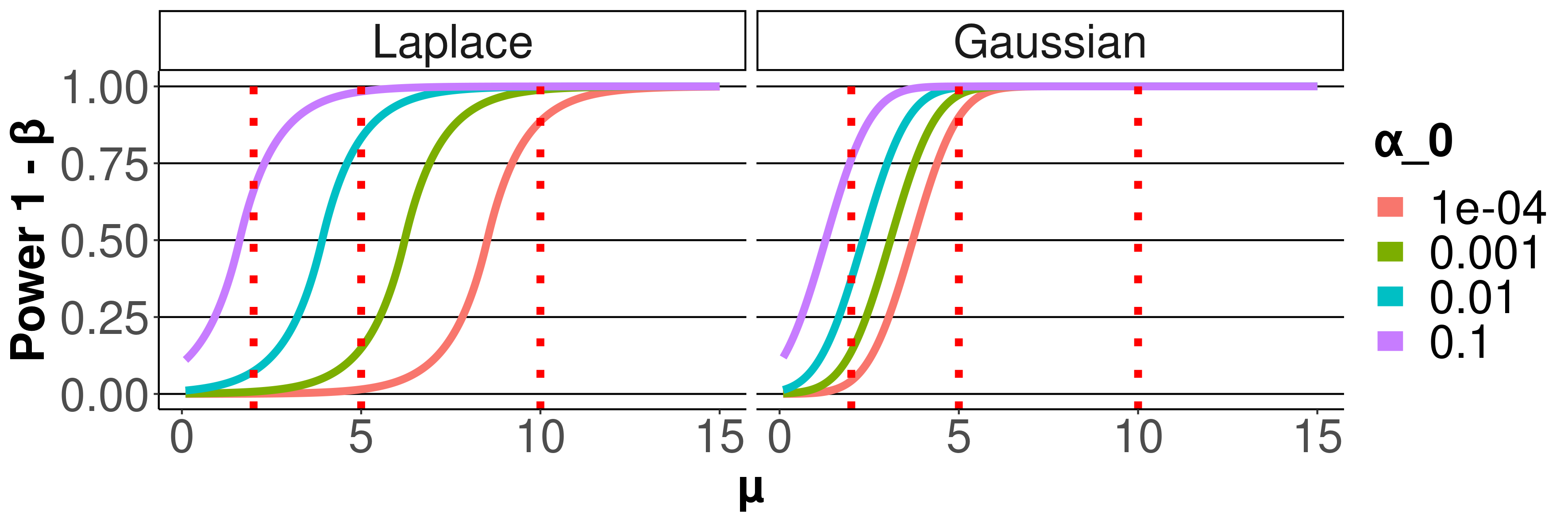}
\caption{Statistical power of the best achievable membership attack on queries protected with the Laplace (left) and Gaussian (right) mechanisms as function of the privacy-loss parameter $\mu$ for multiple choices of false positive rate $\alpha_0$.}
\label{fig: absolute budget}
\end{figure}

Alternatively, we can frame the analysis in terms of test power by selecting the privacy-loss parameter $\mu$ such that the power $1 - \beta(\alpha_0)$ remains below a chosen upper bound, e.g., $0.8$.
This means that, for an attack conducted at a confidence level $\alpha_0$, there remains a probability of at least $\beta$ that true memberships go undetected.
Figure \ref{fig: absolute budget} illustrates the upper bounds on test power across various values of $\alpha_0$ as a function of $\mu$ for both the Laplace and Gaussian mechanisms. The figure confirms that while the Gaussian mechanism  does not offer reasonable guarantees for $\mu \geq 5$, the Laplace mechanism consistently provides stronger protection, even for relatively high values of $\mu$.

\section{Reliable data analysis under differential privacy: An illustrative case study}\label{sec: utility quantification}
From a privacy perspective, data are typically considered fixed in the sense that they represent an underlying ground truth. This is natural, as data protection regulations apply exclusively to collected data, protecting the information as it is measured at the time of collection.
However, in practice, data often suffer from multiple sources of uncertainty, including measurement errors, recording mistakes, and sampling variability. These sources reflect the fact that data represent an imperfect, and often incomplete, snapshot of a world that is itself continuously changing.
%Drawing reliable conclusions from data subject to randomness is the central concern of statistical inference.

In this context, DP can be interpreted as an additional source of uncertainty that must be rigorously analyzed to assess whether valid conclusions can still be drawn. Unlike other forms of variability, the perturbation introduced by DP is precisely specified, which greatly simplifies its mathematical treatment. Moreover, DP guarantees remain valid even when the protection mechanism and its parameter values are disclosed transparently, enabling stakeholders to conduct sound statistical inference on protected algorithmic outputs without compromising privacy.
Viewing DP-induced perturbations through the lens of statistical uncertainty naturally places them within the broader framework of uncertainty propagation: To ensure reliable statistical inference, all sources of uncertainty must be explicitly accounted for and propagated through to the final analytical results. The study of such propagation is a well-established topic in statistics, commonly referred to as measurement error or observational error modeling \citep{Buonaccorsi2010}.

Statistical inference begins with the specification of a parametric model for the data-generating, or data-collection, process.
For instance, a commonly used model assumes additive Gaussian noise:
\begin{equation}\label{eq: data model}
X_i \sim m_i + N(0,\sigma^2), \quad i = 1,\dots, n,  
\end{equation}
where $m_i \in \mathbb{R}$ represents the unobserved ground truth and $\sigma > 0 $ denotes the standard deviation.
Typically, measurement error is modeled directly on the observations, as in \eqref{eq: data model}, but alternative models, such as those accounting for sampling variability, introduce different forms of randomness \citep{Till2006}.

Hypothesis testing provides a principled framework for drawing reliable conclusions from uncertain measurements $D = \{x_1, \dots, x_n\}$, modeled as realizations of random variables $\{X_1,\dots, X_n\}$.
For instance, one may assess whether the result of a query, e.g., a sample mean, is significantly greater than a fixed value $m_0 \in \mathcal{R}$, with a common choice being $m_0 = 0$, by considering the following pair of hypotheses:
\begin{equation}\label{eq: hyp mean}
    \begin{array}{ll}
         H_0: & q(D) = m_0,  \\
         H_1: & q(D) > m_0.
    \end{array}
\end{equation}
Legitimate rejection of $H_0$ in favor of $H_1$ can be analyzed using the framework introduced in Section \ref{sec: f-dp}, with the link between the confidence level $\alpha$ and the power $1-\beta$ conveniently summarized by a ROC curve.

How multiple sources of uncertainty impact the reliability of a conclusion is highly context-dependent, depending on the nature of the data, the data collection process, and how the data are transformed and used. For this reason, modeling uncertainty propagation must be tailored to the specificities of each data analysis. For illustration, we focus on a common setting in which the data are subject to Gaussian noise, as in \eqref{eq: data model}, and the query of interest is the sample mean:
$$
q(D) = \frac{1}{n} \sum_{i = 1}^n x_i \sim N\left(\bar{m},\frac{\sigma^2}{n}\right).
$$
where $\bar{m}$ is the unobserved true population mean.
Although this scenario is simple, it is broadly relevant: by the Central Limit Theorem, the sample mean is approximately normally distributed even when the underlying data are not, provided $n$ is sufficiently large, e.g., $n \geq 12$, highlighting the wide applicability of this setting.
Also, many statistical analyses ultimately reduce to testing whether a group mean takes a particular value. In health research, for example, this includes assessing whether the average blood pressure in a population exceeds a clinical threshold, or whether the mean treatment effect differs from zero in a randomized controlled trial. In economics, similar tests arise when evaluating whether the average income of a group differs from a policy target, or whether a macroeconomic indicator such as mean unemployment duration has changed following an intervention. In the social sciences, researchers frequently test whether the average response to a survey question differs from a neutral baseline, or whether the mean level of educational attainment in a population meets a predefined benchmark.
Clearly, all of these highly relevant questions require the analysis of personal and sensitive data, the confidentiality of which must be protected.

We now consider a setting in which the query result is protected via an $f$-DP mechanism, and we adapt both the model and the corresponding hypotheses accordingly. For illustration, we consider the Gaussian mechanism, which yields simple expressions:
$$
Q(D) = q(D) + N = N\left(\bar{m},\frac{\sigma^2}{n} + \frac{\Delta q^2}{\mu^2}\right),
$$
where $\Delta q = \sup_{i,i' \in [1,n]} |x_i -x_{i'}| / n$.
Since the added noise has zero mean, the hypotheses in \eqref{eq: hyp mean} remain unchanged, only the test power is affected by the DP-mechanism.

Hypothesis testing under this model can be performed using a Z-test \citep{Fisher1925}. Since the alternative hypothesis $H_1$ does not specify a unique parameter value, the test power cannot be summarized by a single ROC curve. In this setting, a classical tool is the power function, which describes the test’s power at a fixed significance level $\alpha_0$, commonly chosen as $\alpha_0 = 0.05$ or $\alpha_0 = 0.01$, as a function of a candidate alternative means $m$:
\begin{equation}\label{eq: type II under DP}
    1 - \beta_{\alpha_0}(m) = 1 - F\left\{F^{-1}(1 - \alpha_0)  - \frac{m}{\sigma_{n}} \right\},
\end{equation}
where $\sigma_n^2 = \frac{\sigma^2}{n} + \frac{\Delta q^2 }{\mu^2}$ and $F$ denotes the cumulative distribution function of a standard normal distribution.

\begin{figure}%[tbhp]
\centering
\begin{tabular}{c}
     \includegraphics[width=\linewidth]{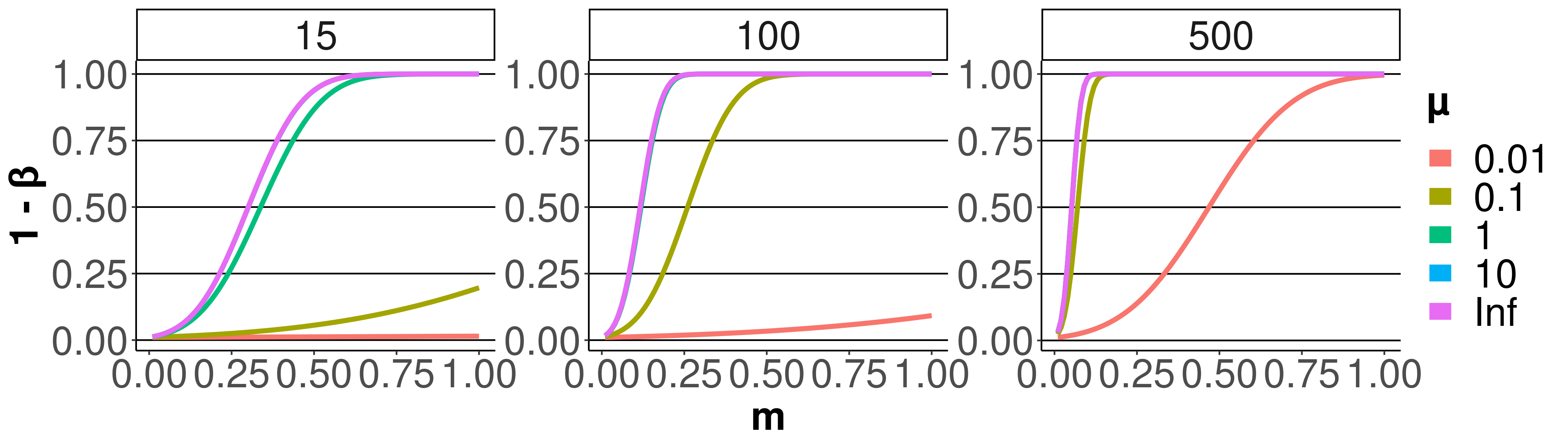} \\\includegraphics[width=\linewidth]{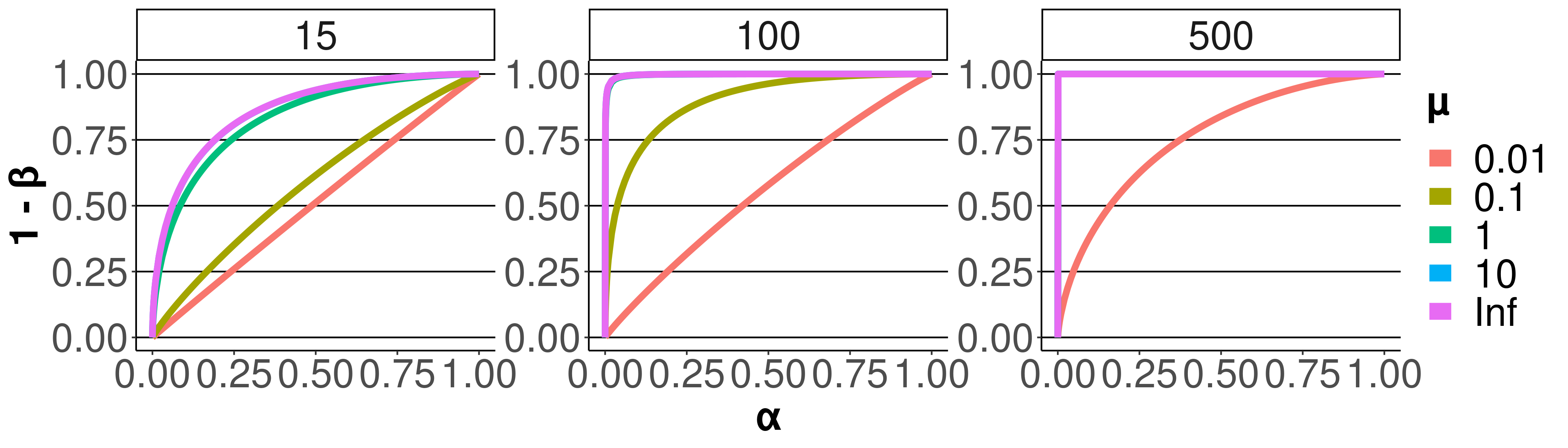}
\end{tabular}
\caption{Top: Power function of a Z-test at a confidence level $\alpha_0 = 0.01$ for a mean query protected with a Gaussian mechanism over data $x_i \in [0,1]$ suffering from Gaussian noise with $\sigma = 0.25$. Bottom: ROC-curve of a Z-test with alternative mean $m = 0.2$. for a mean query protected with a Gaussian mechanism. The privacy loss parameter $\mu = \infty$ correspond to the release of a non-protected query.}
\label{fig: mean diff hypothesis}
\end{figure}

Figure \ref{fig: mean diff hypothesis} shows, for binary data $X \in \{0,1\}$, the power function at confidence level $\alpha_0 = 0.01$, and the corresponding ROC curves for $m = 0.2$, across various sample sizes $n = 15,100,500$. Notably, statistical power remains virtually unaffected for privacy parameters $\mu  \geq 1$, across all sample sizes. For largest datasets ($n=500$), we observe a substantial loss in power only when $\mu = 0.01$.
This demonstrates that strong privacy protection under $f$-DP can be achieved without significantly compromising statistical utility.
%In this example, the strength of the privacy guarantee even improves with increasing dataset size.

The above analysis can be extended to a wide variety of statistical models, tests, and DP mechanisms. This includes, for example, testing parameter equality across subgroups, or assessing the significance of coefficients in linear regression models. Even when uncertainty propagation does not yield closed-form solutions, the statistical power of a given test can still accurately be estimated via simulation \citep[Section 2.4.2]{Meier2022}.
Treating DP-induced perturbation as an additional, well-characterized source of uncertainty to be propagated through statistical analysis is not itself new: it dates back to \citet{vu2009differential}, who first proposed accounting for DP noise when analyzing clinical trial data, and was subsequently extended through analytical approximations to the sampling distributions of common estimators \citep{wang2018statistical}. More recently, simulation-based and Bayesian approaches have been developed to conduct valid inference directly from privatized data, including parametric bootstrap methods \citep{ferrando2022parametric}, finite-sample simulation-based inference \citep{awan2025simulation}, data-augmentation MCMC \citep{ju2022data}, and exact inference via approximate computation \citep{gong2022exact}.
Guidance for conducting such analyses while prioritizing the reliability of the analysis is provided in Section \ref{sec: utility first}.

\section{The privacy-utility trade-off in practice}\label{sec: trade-off}
%\section{The privacy-utility trade-off}
Collecting data is not an end in itself; rather, it serves to generate value from the information it contains. However, this value is rarely obtained directly from raw data. Instead, the data are typically transformed into derived or generated data products. Unlocking the benefits of these products often entails sharing them widely, including with actors with potentially malicious intents.
Consequently, these data products must be safeguarded to limit the risk that their dissemination adversely affects the individuals providing the underlying data.
 
%Historically, data protection has relied on anonymization, ensuring that individuals cannot be re-identified from the disseminated data. However, absolute anonymization, which guarantees irreversibility, has proven to be a broken promise: anonymization techniques are today known to be vulnerable to membership inference \cite[e.g.,][]{homer2008} , reconstruction \cite[e.g.,][]{abowd2023}, and linkage attacks \cite[e.g.,][]{sweeney2000}, depending on the nature of the data product. It is now imperative to shift our mindset and acknowledge that the risk of disclosure when sharing a data product is never zero.
%Communicating this reality transparently is essential.

%Data products are created with the expectation of delivering value or benefits but their dissemination expose the original data at risk disclosure which can have harmful consequences.
The risk of data disclosure is zero only if no data products are released. Any dissemination, even in aggregated or otherwise derived form, creates some potential for information leakage, so absolute anonymization is possible only in the absence of release. Consequently, deriving value from data requires balancing privacy and utility by weighing disclosure risk against expected benefits.
To navigate this trade-off effectively, it is crucial to clearly define
\begin{itemize}
    \item the purpose of data usage,often referred to as a project or data analysis, along with the expected benefits,
    \item the nature of the data: the more sensitive the data, the lower the acceptable risk, which makes it increasingly challenging to justify the use and dissemination of the results, especially if they are susceptible to a high likelihood of disclosure,
    \item the data recipients of the data product(s), meaning those who will access and use the results.
\end{itemize}
Governance frameworks, such as the Five Safes \citep{desai2016} or the Fair Information Practice Principles (FIPPs) \citep{FIPP_2008}, help systematically evaluate these dimensions.

Building upon this context, there are two main paradigms for managing the privacy-utility trade-off:
\begin{itemize}
    \item Privacy-first: This approach starts by defining the maximum acceptable risk, considering the project’s purpose, data sensitivity, and data recipients. The acceptable risk level is then translated into a privacy guarantee through the selection of appropriate privacy-loss parameters. Then, the resulting protected data product is analyzed to determine whether its utility is sufficient for drawing reliable conclusions. If the utility is too low to meet the project’s objectives, the relevance of the data product creation is questioned: the desired benefits are tied to a not acceptable level of risk of disclosure.
    \item Utility-first: Here, the process begins by assessing the required utility of the protected data product and determining the maximum allowable perturbation that guarantee sufficient reliability. The chosen level of perturbation is then mapped to a privacy guarantee, defining the minimal associated disclosure risk to enjoy the product's benefits. Accepting the project under this approach means acknowledging that utility outweigh the necessary privacy risks.
\end{itemize}

The results of Sections \ref{sec: privacy garantee} and \ref{sec: utility quantification} provide a direct connection between an interpretable notion of disclosure risk, namely the relative disclosure risk, and the ability to draw reliable conclusions from the protected data product. This framework provides a structured way to evaluate and communicate the impact of privacy protection on data utility as well as select appropriate values for the privacy-loss parameters.

\subsection{Privacy-first: choosing a DP-Mechanism and its parameters}
Table \ref{tab: privacy guarantees} provides an overview of the guarantees and potential failure modes allowing randomized mechanisms to be classified into three families whose characteristics, even though mathematical, are intended to also be interpretable by non-experts.
\begin{table}[t!]
\centering
\begin{tabular}{ccrr}
$f(0)$ & $\rho$ &  (Relative) Disclosure Risk & Failure Type \\
\midrule
$1$ & $< \infty$ &  Strictly Bounded by $\rho$ & None \\
$1$ & $\infty$ & Bounded by $\rho_{\alpha_0}$ at level $\alpha_{0}$  & Graceful \\
$< 1$ & $\infty$ &  Blatant disclosure & Catastrophic\\
\bottomrule
\end{tabular}
\caption{Categories of randomized mechanisms and their corresponding guarantees as a function of their maximum relative disclosure risk and the properties of their trade-off functions.}
\label{tab: privacy guarantees}
%\addtabletext{nomenclature for the TSs refers to the numbered species in the table.}
\end{table}
%This classification provides a comprehensive overview of the spectrum of privacy guarantees offered by different classes of mechanisms. 
Under a privacy-first paradigm, this classification guides the selection of an appropriate mechanism and its associated privacy-loss parameter $\mu$, which can be summarized by the following set of questions:
\begin{enumerate}
    \item Are blatant disclosure acceptable?
    \begin{itemize}
    \item "Yes" $\Leftrightarrow$ $f(0) < 1$: All DP mechanisms, including those susceptible to catastrophic failure, may be considered.
    \item "No" $\Leftrightarrow$ $f(0) = 1$: Mechanisms prone to catastrophic disclosure must be excluded from consideration. 
    \end{itemize}
    Determining the answer requires evaluating whether the data recipients of differentially private data products, ranging from trusted collaborators under confidentiality agreements (e.g., public servants) to the general public (e.g., scientific publications or open government data), are likely to attempt membership inference attacks or have the technical means to do so. Dissemination of products susceptible to obvious disclosure may be acceptable in cases where the circle of data recipients is controlled. A good example is the distribution of fine-grained counts of viral diseases cases: while public release of such data could pose a high risk of disclosure, the risk is substantially lower when dissemination is limited to local authorities using those numbers for public policy purposes only. Finally, the nature of the data also influences directly the answer:for instance, the acceptability of disclosure risk is lower for health data than for information such as cinema attendance.
    \item Is it acceptable for membership attacks to achieve arbitrarily high confidence?
    \begin{itemize}
     \item "No" $\Leftrightarrow$ $f(0) = 1$ and $\rho < \infty$: Only mechanisms with strictly bounded relative disclosure risk should be selected.
     \item "Yes" $\Leftrightarrow$ $f(0) = 1$ but $\rho = \infty$: Mechanisms with graceful failure may also be included.
    \end{itemize}
    In this case, although blatant disclosure is avoided, it may still be possible in rare instances for an attacker to achieve a level of confidence approaching certainty. For instance, by accepting a very high likelihood of failing to detect an individual who is actually in the dataset, the attacker can ensure that a small number of successful attacks yield near-certain confidence. This strategy is relevant to an attacker when targeting a large pool of candidates: although many true members may be missed, any successful detections are definitive and actionable. If such a scenario is considered unlikely in the context of dissemination, then arbitrarily high relative risk may be deemed acceptable.
    \item What is the maximum tolerable disclosure risk for the intended application? Two complementary approaches can be used:
    \begin{itemize}
        \item Maximum relative disclosure risk $R > 0$: 
        Choose the privacy-loss parameter $\mu$ such that $\rho = R$, or $\rho_{\alpha_0} = R$, depending on whether the mechanism has bounded or unbounded relative disclosure risk. This limits the factor by which the confidence over a target's membership can be increased due to the disseminated queries.
        \item Maximum attack power $1 - \beta_0$ at confidence level $\alpha_0$:
        Select $\mu$ such that $f_\mu(\alpha_0) = \beta_0$. This frames disclosure risk in terms of the statistical power of the strongest possible membership attack at a specified false positive rate $\alpha_0$.
    \end{itemize}
    Both methods are formally equivalent, differing only in their interpretation. Relative disclosure risk is often more intuitive for non-technical stakeholders, as it avoids the need to explain concepts such as Type I and Type II errors.
    In general, the selection of the numerical value of $R$ should focus on the order of magnitude rather than exact values.
    Evaluating concrete disclosure scenarios for realistic values of $\alpha$, such as  $\alpha = 0.01$, can help refine the choice of privacy-loss parameters within the selected order of magnitude. This requires additional assumptions regarding the technical capabilities of plausible attacker candidates, as well as the nature and amount of information available to them. The purpose of this exercise is not to discuss the formal privacy guarantee itself but to present and communicate the practical impact of selecting particular values for the privacy-loss parameters in an accessible way.
\end{enumerate}

This decision framework facilitates the principled selection of both the randomized mechanism and its privacy parameters. In the absence of additional constraints, the mechanism offering the strongest privacy guarantee should be preferred.
However, as discussed at the beginning of this Section, stronger privacy typically entails reduced utility. Hence, a careful analysis must assess the extent to which the randomized query output $q$ remains usable. Such an evaluation is essential to determine whether the differentially private result can support the intended analytical or operational goals with sufficient reliability.

\subsection{Utility-first: linking the privacy-loss parameters with statistical significance}\label{sec: utility first}
Under the utility-first paradigm, the objective is to determine the maximum level of privacy protection that can be offered while maintaining a predefined level of utility. This analysis must be conducted carefully and requires distinguishing between two primary operational settings:
\begin{itemize}
    \item Central curator: In this model, the data owner is responsible for authorizing, generating, and disseminating the query result. Authorized personnel operate under legal and organizational constraints that grant access to the raw data.% allowing them to develop and execute the utility-privacy analysis using the actual data.
    \item ``Eyes-off'' users: Here, the data product is created by the data recipients themselves who interact with a system or platform, such as \cite{Aymon2024}, that permits the execution of DP-protected algorithms remotely, without direct access to the underlying data.
\end{itemize}
When analyzing the reliability of data analyses, these two configurations can be mapped to two types of statistical inference:
\begin{itemize}
    \item Observational studies analyze pre-existing data and allow for flexible, exploratory approaches, where the analysis plan can evolve during the process.
    \item Experimental studies aim to test causal hypotheses and require the pre-specification of the analysis plan to ensures that the data collection strategy yields sufficient statistical power for reliable inference. Because the analysis is conducted prior to data collection, the plan must be devised without any feedback from the data.
\end{itemize}
Although utility-privacy trade-offs are often considered after data have already been collected, the utility analysis should ideally be conducted as if the data were not yet available, mirroring the methodology of experimental studies.
This approach serves two critical purposes. First, it ensures that the utility analysis introduces no additional disclosure risk, as it is performed independently of the data. Second, it mitigates selection and confirmation biases, whereby analysis decisions are unduly influenced by characteristics observed in the data. Such biases can undermine reproducibility and, ultimately, the reliability of results.
%Therefore, the evaluation of the utility of a DP-protected data product should ideally be conducted without direct access to the raw data by leveraging auxiliary metadata information and accounting all potential sources of uncertainty that could affect the robustness of the conclusions.

In practice, utility analysis without access to the data requires the selection of plausible values for model parameters such as $\sigma$ in \eqref{eq: data model}.
%In central curator scenarios, it is relatively safe to rely on empirical estimates; however,
We recommend justifying their values using characteristics of the data that are considered public, or any piece of information that could be available before collection.
For instance, a variance of $0.25$ for a Bernoulli-distributed binary variable is a known upper bound. Adopting a conservative, i.e., worst-case, estimate ensures that the conclusions remain valid even if the true variance is smaller.

A similar consideration applies to dataset size, which plays a pivotal role in determining the statistical power of many tests and is also a required parameter for some DP mechanisms. In many cases, dataset size is treated as public information. If this is not the case, it is reasonable to allocate a small portion of the privacy-loss parameter to estimate the dataset size, or any other quantity relevant to the utility analysis, and then use a conservative lower bound of these quantities for the utility analysis.
For this purpose, synthetic datasets generated using differentially private generators can play a pivotal role. Algorithms such as MST \citep{McKenna2021} can be used to produce synthetic data that reproduce counts, means, and variances across all combinations of categorical variables while limiting disclosure risk.
These synthetic datasets are not intended for drawing reliable inferential conclusions; rather, they serve the sole purpose of enabling data recipients who do not have direct access to the original data to conduct their own utility analyses, following the modeling strategy described in Section \ref{sec: utility quantification}.
These datasets can be generated in advance by the central curator and, provided that the privacy-loss parameter used in their generation is sufficiently small, made publicly available to facilitate feasibility studies and to identify the minimal privacy-loss parameter required for the realization of a given project.

Continuing the example from Section \ref{sec: utility quantification}, we can compute the minimum privacy-loss parameter $\mu$ required to ensure that reliability degrades by no more than, for instance, 
$1\%$ after the application of the DP mechanism.
Specifically, this requirement implies that the type II error $\beta$ at level $\alpha$, with $\alpha = 0.05$ being a very common standard, does not increase by more than $1\%$, i.e.,
$$
\inf_{\mu > 0} \left\{\left|\frac{\beta_\mu(\alpha) - \beta(\alpha)}{\beta(\alpha)}\right| \leq 0.01 \right\}.
$$
Using Equation \eqref{eq: type II under DP}, this yields minimum privacy parameters of $\mu_{\min} = 7.9, 0.28, 0.048$ for $n = 15,100,500$, respectively, in order to remain within $1\%$ of the unprotected performance.
Alternatively, $\mu$ can be chosen as the smallest value that ensures a test power of at least $0.8$, i.e., $\beta_\mu(\alpha) \geq 0.8$, which is a widely accepted standard in the health sciences.
Once the minimum privacy-loss parameter $\mu$ is determined, the results of Section \ref{sec: privacy garantee} allow the corresponding privacy guarantee to be expressed in an interpretable form, facilitating a principled assessment of the privacy–utility trade-off.

While the example presented in Section \ref{sec: utility quantification} illustrates a wide range of applications, it remains overly simplistic, as real-world data analyses rarely focus on a single hypothesis.
In practice, multiple hypotheses or statistical tests are often conducted, which increases the probability of at least one false positive and can lead to misleading conclusions. It is therefore essential to incorporate appropriate correction methods into the statistical analysis plan, using techniques such as the Bonferroni correction \citep{Miller1981} or False Discovery Rate control procedures \citep{Benjamini1995}.

\section*{Discussion} 
In this paper, we have demonstrated how the privacy–utility trade-off can be rigorously balanced using differential privacy and statistical hypothesis testing. Specifically, framing membership inference attacks within the $f$-differential privacy framework directly links privacy to a classical statistical inference problem: drawing reliable conclusions from uncertain data.
This approach enables a concrete and interpretable understanding of the privacy-loss parameters in terms of blatant disclosure, relative disclosure risk, and statistical power.
This work represents an initial effort to bridge statistical methodology and DP. We emphasize that the formal impact of noise perturbation can be rigorously analyzed using tools from statistical inference. The illustrative yet widely applicable case of Gaussian uncertainty combined with the Gaussian mechanism is extendable to other settings, such as linear regression \cite[Section 2]{Meier2022}, and can be integrated with advanced tools for multiple hypothesis testing \citep{Miller1981, Bretz2011}. As a result, the reliability of statistical conclusions under DP perturbations can be formally quantified in diverse contexts, offering a broad avenue for utility evaluation.

From a privacy standpoint, we observe that membership inference attacks tend to yield conservative privacy parameter allocations. This outcome stems from relatively strong, albeit legitimate, assumptions about the adversary’s knowledge and capabilities. Extending the analysis to encompass reconstruction attacks could potentially yield less restrictive parameter constraints, while preserving or even enhancing their practical relevance. However, this extension would necessitate a more nuanced treatment of hypothesis testing, particularly to address the multiple comparisons inherent in optimization routines. In such cases, the privacy-loss parameters could be linked to the expected success rate of reconstruction attacks, thereby providing even more actionable guidance in choosing and communicating privacy parameters.

%Moreover, our analysis adopts a conservative perspective on the information available to attackers prior to an attack, considering all possible prior distributions. Recent work by the U.S. Census Bureau \citep{abowd2023} has explored a similar analysis constrained to specific families of probabilistic forecast baselines. Their objective is to assess how much these baselines can be improved through the use of disseminated data products. A comparable strategy could be combined with our methodology for concrete use cases, such as the Census tabular data releases, where reasonable and realistic prior baselines can be clearly identified.

This work lays the foundation for a comprehensive governance framework to guide the creation and dissemination of data products under two paradigms. In a utility-first approach, the process begins with a systematic evaluation of the expected benefits of the algorithmic outputs, with the aim of determining the minimum disclosure risk that must be tolerated to realize those benefits.
Under the privacy-first paradigm, by contrast, the entity responsible for safeguarding the outputs treats the selection of privacy-loss parameters as a policy decision: how much should the uncertainty faced by potential attackers be reduced, and thus how much easier should it be for them to infer sensitive information? This choice, in turn, determines the level of reliability analysts can expect from the results.
%under either privacy-first or utility-first paradigms.
%Under the privacy-first paradigm, the entity responsible for safeguarding algorithmic outputs must treat the choice of a privacy-loss parameters as a policy decision: to what extent is it acceptable to reduce the uncertainty faced by potential attackers, i.e., how much easier should it be for them to infer sensitive information? This decision, in turn, determines the level of reliability an analyst can expect from the results.
%In contrast, utility-first strategies start with the systematic evaluation of the expected benefits provided by the algorithmic outputs, in order to determine the minimal risk that must be tolerated to realize those benefits.
%The framework enables consideration of the underlying data, the intended data recipients, and the characteristics of the data product, while delivering a rigorous mathematical quantification of disclosure risk and utility degradation.
%This work offers a new perspective on the guarantees provided by differential privacy: namely, that
By modeling privacy attacks as decision-making under uncertainty, DP’s guarantee can be reinterpreted as ensuring the unreliability of membership inference attacks while bounding the maximum attainable gain in confidence.
From a legal standpoint, anonymization has so far required that data be irreversibly transformed such that individual identification is no longer possible, which cannot technically be guaranteed.
However, one might ask legal scholars: is the unreliability guarantee sufficient to claim the irreversibility and meet the threshold of anonymity? If not, is there a quantifiable reliability threshold below which data can be considered anonymized in broad generality? Has the very concept of anonymization become obsolete in light of recent advances in methodology and technology? In contexts where analyses are performed on inherently uncertain data, individuals may still be exposed to harm, even if disclosure is not certain. These remain open questions, reflecting the ongoing challenges to protect privacy.

Finally, by framing the act of privacy-utility balance through the lens of statistical inference, we will also be able to address broader issues of inferential privacy, that is, the protection of individual-level information that may be estimated or inferred from a data product, regardless of whether the individual's data contributed to its creation. Sometimes termed generalization or inference risk, this form of disclosure is intrinsically uncertain, requiring statistical tools for robust risk quantification and giving rise to a line of research in which statistical inference is used not to enhance accuracy, but to protect individuals from harm.

This discussion underscores the inherently interdisciplinary nature of the field. Legal scholars, policymakers, and technical experts in privacy and data science must collaborate to establish a shared understanding. Only through such cooperation can we develop a pragmatic and resilient framework for data privacy, one that reflects the complexities of today’s digital and technological landscape.

\section{Acknowledgements}
The author acknowledges the indispensable support of the Federal Statistical Office. The author is also grateful to the OpenDP fellowship program for fostering valuable discussions and the exchange of ideas. Finally, the author thanks Christine Choirat for her insightful discussions and excellent feedback, which significantly improved the quality of this paper.

% Bibliography
\bibliography{lomas}

@book{Gelman2020,
   author = {Andrew Gelman and John B Carlin and Hal S Stern and David B Dunson and Aki Vehtari and Donald B. Rubin},
   city = {New York},
   publisher = {Chapman and Hall/CRC},
   title = {Bayesian Data Analysis},
   year = {2020}
}

@unpublished{sweeney2000,
   author = {Latanya Sweeney},
   city = {Pittsburgh},
   institution = {Carnegie Mellon University},
   publisher = {Sweeney},
   title = {Simple Demographics Often Identify People Uniquely},
   year = {2000}
}

@article{Dwork2014,
   author = {Cynthia Dwork and Aaron Roth},
   doi = {10.1561/0400000042},
   issn = {15513068},
   issue = {3-4},
   journal = {Foundations and Trends in Theoretical Computer Science},
   pages = {211-407},
   publisher = {Now Publishers Inc},
   title = {The Algorithmic Foundations of Differential Privacy},
   volume = {9},
   year = {2014}
}

@article{Dong2022,
   author = {Jinshuo Dong and Aaron Roth and Weijie J. Su},
   doi = {10.1111/rssb.12454},
   issn = {14679868},
   issue = {1},
   journal = {Journal of the Royal Statistical Society. Series B: Statistical Methodology},
   month = {2},
   pages = {3-37},
   publisher = {John Wiley and Sons Inc},
   title = {Gaussian differential privacy},
   volume = {84},
   year = {2022}
}

@inproceedings{dwork2006,
   author = {Cynthia Dwork},
   city = {Heidelberg},
   booktitle = {International colloquium on automata, languages, and programming},
   pages = {1-12},
   publisher = {Springer Berlin},
   title = {Differential Privacy},
   year = {2006}
}

@unpublished{desai2016,
   author = {Tanvi Desai and Felix Ritchie and Richard Welpton},
   city = {Bristol},
   institution = {University of the West of England},
   title = {Five Safes: designing data access for research},
   url = {https://www2.uwe.ac.uk/faculties/BBS/Documents/1601.pdf},
   year = {2016}
}

@article{homer2008,
   author = {Nils Homer and Szabolcs Szelinger and Margot Redman and David Duggan and Waibhav Tembe and Jill Muehling and John V. Pearson and Dietrich A. Stephan and Stanley F. Nelson and David W. Craig},
   doi = {10.1371/journal.pgen.1000167},
   issn = {15537390},
   issue = {8},
   journal = {PLoS Genetics},
   month = {8},
   pmid = {18769715},
   title = {Resolving individuals contributing trace amounts of DNA to highly complex mixtures using high-density SNP genotyping microarrays},
   volume = {4},
   year = {2008}
}

@article{abowd2023,
   author = {John M. Abowd and Tamara Adams and Robert Ashmead and David Darais and Sourya Dey and Simson L. Garfinkel and Nathan Goldschlag and Daniel Kifer and Philip Leclerc and Ethan Lew and Scott Moore and Rolando A. Rodríguez and Ramy N. Tadros and Lars Vilhuber},
   journal = {arXiv:2312.11283},
   month = {12},
   title = {The 2010 Census Confidentiality Protections Failed, Here's How and Why},
   url = {http://arxiv.org/abs/2312.11283},
   year = {2023}
}

@article{Kifer2022,
   author = {Daniel Kifer and John M. Abowd and Robert Ashmead and Ryan Cumings-Menon and Philip Leclerc and Ashwin Machanavajjhala and William Sexton and Pavel Zhuravlev},
   journal = {arXiv:2209.03310},
   month = {9},
   title = {Bayesian and Frequentist Semantics for Common Variations of Differential Privacy: Applications to the 2020 Census},
   url = {http://arxiv.org/abs/2209.03310},
   year = {2022}
}

@article{Dwork2017,
   author = {Cynthia Dwork and Adam Smith and Thomas Steinke and Jonathan Ullman},
   doi = {10.1146/annurev-statistics-060116-054123},
   issn = {2326831X},
   journal = {Annual Review of Statistics and Its Application},
   month = {3},
   pages = {61-84},
   publisher = {Annual Reviews Inc.},
   title = {Exposed! A survey of attacks on private data},
   volume = {4},
   year = {2017}
}

@article{Dwork2019,
   author = {Cynthia Dwork and Nitin Kohli and Deirdre Mulligan},
   doi = {10.29012/jpc.689},
   issn = {25758527},
   issue = {2},
   journal = {Journal of Privacy and Confidentiality},
   month = {10},
   pages = {1-22},
   publisher = {Cornell University},
   title = {Differential Privacy in Practice: Expose your Epsilons!},
   volume = {9},
   year = {2019}
}

@book{Marcum1947,
   author = {JI. Marcum},
   city = {Santa Monica, CA},
   publisher = {RAND Corporation},
   title = {A Statistical Theory of Target Detection by Pulsed Radar},
   year = {1947}
}

@book{Lehmann2022,
   author = {E. L. Lehmann and Joseph P. Romano},
   city = {Cham},
   editor = {Springer Nature Switzerland AG},
   title = {Testing Statistical Hypotheses},
   year = {2022}
}

@techReport{FIPP_2008,
   author = {U.S. Department of Homeland Security},
   city = {Washington},
   month = {12},
   title = {Privacy Policy: Fair Information Practice Principles},
   url = {http://aspe.hhs.gov/DATACNCL/1973privacy/tocprefacemembers.htm},
   year = {2008}
}

@book{Buonaccorsi2010,
   author = {John P. Buonaccorsi},
   city = {New York},
   publisher = {Chapman and Hall/CRC},
   title = {Measurement Error - Models, Methods, and Applications },
   year = {2010}
}

@book{Fisher1925,
   author = {R.A. Fisher},
   city = {Edinburgh},
   publisher = {Oliver \& Boyd},
   title = { Statistical Methods for Research Workers},
   year = {1925}
}

@book{Meier2022,
   author = {Lukas Meier},
   publisher = {CRC Press},
   title = {ANOVA and Mixed Models - A Short Introduction Using R},
   year = {2022}
}

@book{Miller1981,
   author = {Rupert G. Miller},
   city = {New York, NY},
   publisher = {Springer},
   title = {Simultaneous Statistical Inference Book},
   year = {1981}
}

@book{Bretz2011,
   author = {Frank Bretz and Torsten Hothorn and Peter Westfall},
   city = {New York},
   publisher = {Chapman and Hall/CRC},
   title = {Multiple Comparisons Using R},
   year = {2011}
}

@article{Benjamini1995,
   author = {Yoav Benjamini and Yosef Hochberg},
   issue = {1},
   journal = {Journal of the Royal Statistical Society. Series B (Methodological)},
   pages = {289-300},
   title = {Controlling the False Discovery Rate: A Practical and Powerful Approach to Multiple Testing},
   volume = {57},
   year = {1995}
}

@article{Su2025,
   author = {Weijie J Su},
   doi = {10.1146/annurev-statistics-112723},
   isbn = {202509:17:29},
   journal = {Annual Review of Statistics and Its Application},
   pages = {157-175},
   title = {Annual Review of Statistics and Its Application A Statistical Viewpoint on Differential Privacy: Hypothesis Testing, Representation, and Blackwell's Theorem},
   volume = {12},
   url = {https://doi.org/10.1146/annurev-statistics-112723-},
   year = {2025}
}

@article{Warren1980,
   author = {Samuel D. Warren and Louis D. Brandeis},
   issue = {5},
   journal = {Harvard Law Review},
   pages = {193-220},
   title = {The right to privacy},
   volume = {4},
   year = {1890}
}

@article{Aymon2024,
   author = {Damien Aymon and Dan-Thuy Lam and Lancelot Marti and Pauline Maury-Laribière and Christine Choirat and Raphaël de Fondeville},
   journal = {arXiv:2406.17087},
   month = {6},
   title = {Lomas: A Platform for Confidential Analysis of Private Data},
   url = {http://arxiv.org/abs/2406.17087},
   year = {2024}
}

@book{Till2006,
   author = {Yves Tillé},
   city = {Berlin, Heidelberg},
   publisher = {Springer},
   title = {Sampling algorithms},
   year = {2006}
}

@article{McKenna2021,
   author = {Ryan McKenna and Gerome Miklau and Daniel Sheldon},
   issue = {3},
   journal = {Journal of Privacy and Confidentiality},
   title = {Winning the NIST Contest: A scalable and general approach to differentially private synthetic data},
   volume = {11},
   year = {2021}
}

@article{sweeney2002,
  author  = {Latanya Sweeney},
  title   = {k-Anonymity: A Model for Protecting Privacy},
  journal = {International Journal of Uncertainty, Fuzziness and Knowledge-Based Systems},
  volume  = {10},
  number  = {5},
  pages   = {557--570},
  year    = {2002},
  doi     ={10.1142/S0218488502001648}
}

@inproceedings{kifer2011nofreelunch,
  author    = {Kifer, Daniel and Machanavajjhala, Ashwin},
  title     = {No Free Lunch in Data Privacy},
  booktitle = {Proceedings of the 2011 ACM SIGMOD International Conference on Management of Data},
  series    = {SIGMOD '11},
  pages     = {193--204},
  year      = {2011},
  publisher = {Association for Computing Machinery},
  address   = {New York, NY, USA},
  doi       = {10.1145/1989323.1989345}
}

@article{kasiviswanathan2014semantics,
  author  = {Kasiviswanathan, Shiva Prasad and Smith, Adam},
  title   = {On the 'Semantics' of Differential Privacy: A {B}ayesian Formulation},
  journal = {Journal of Privacy and Confidentiality},
  volume  = {6},
  number  = {1},
  pages   = {1--16},
  year    = {2014},
  doi     = {10.29012/jpc.v6i1.634}
}

@inproceedings{mcsherry2007mechanism,
  author    = {McSherry, Frank and Talwar, Kunal},
  title     = {Mechanism Design via Differential Privacy},
  booktitle = {Proceedings of the 48th Annual IEEE Symposium on Foundations of Computer Science (FOCS)},
  pages     = {94--103},
  year      = {2007},
  publisher = {IEEE},
  doi       = {10.1109/FOCS.2007.66}
}

@inproceedings{smith2011privacy,
  author    = {Smith, Adam},
  title     = {Privacy-Preserving Statistical Estimation with Optimal Convergence Rates},
  booktitle = {Proceedings of the 43rd Annual ACM Symposium on Theory of Computing (STOC)},
  pages     = {813--822},
  year      = {2011},
  publisher = {Association for Computing Machinery},
  doi       = {10.1145/1993636.1993743}
}

@article{wasserman2010statistical,
  author  = {Wasserman, Larry and Zhou, Shuheng},
  title   = {A Statistical Framework for Differential Privacy},
  journal = {Journal of the American Statistical Association},
  volume  = {105},
  number  = {489},
  pages   = {375--389},
  year    = {2010},
  doi     = {10.1198/jasa.2009.tm08651}
}

@article{ponomareva2023dpfy,
  author  = {Ponomareva, Natalia and Hazimeh, Hussein and Kurakin, Alex and Xu, Zheng and Denison, Carson and McMahan, H. Brendan and Vassilvitskii, Sergei and Chien, Steve and Thakurta, Abhradeep},
  title   = {How to {DP-fy} {ML}: A Practical Guide to Machine Learning with Differential Privacy},
  journal = {Journal of Artificial Intelligence Research},
  volume  = {77},
  pages   = {1113--1201},
  year    = {2023}
}

@article{kazan2023prioritizing,
  author  = {Kazan, Zeki and Reiter, Jerome P.},
  title   = {Prior-itizing Privacy: A {B}ayesian Approach to Setting the Privacy Budget in Differential Privacy},
  journal = {arXiv preprint arXiv:2306.13214},
  year    = {2023}
}

@inproceedings{ligett2017accuracy,
  author    = {Ligett, Katrina and Neel, Seth and Roth, Aaron and Waggoner, Bo and Wu, Z. Steven},
  title     = {Accuracy First: Selecting a Differential Privacy Level for Accuracy-Constrained {ERM}},
  booktitle = {Advances in Neural Information Processing Systems 30 (NeurIPS 2017)},
  year      = {2017}
}

@article{whitehouse2022brownian,
  author  = {Whitehouse, Justin and Ramdas, Aaditya and Rogers, Ryan and Wu, Zhiwei Steven},
  title   = {Brownian Noise Reduction: Maximizing Privacy Subject to Accuracy Constraints},
  journal = {arXiv preprint arXiv:2206.07234},
  year    = {2022}
}

@article{mcclure2012differential,
  author  = {McClure, David and Reiter, Jerome P.},
  title   = {Differential Privacy and Statistical Disclosure Risk Measures: An Investigation with Binary Synthetic Data},
  journal = {Transactions on Data Privacy},
  volume  = {5},
  number  = {3},
  pages   = {535--552},
  year    = {2012}
}

@article{abowd2022topdown,
  author  = {Abowd, John M. and Ashmead, Robert and Cumings-Menon, Ryan and Garfinkel, Simson and Heineck, Micah and Heiss, Christine and Johns, Robert and Kifer, Daniel and Leclerc, Philip and Machanavajjhala, Ashwin and Moran, Brett and Sexton, William and Spence, Matthew and Zhuravlev, Pavel},
  title   = {The 2020 {C}ensus Disclosure Avoidance System {TopDown} Algorithm},
  journal = {Harvard Data Science Review},
  volume  = {Special Issue 2},
  year    = {2022},
  doi     = {10.1162/99608f92.529e3cb9}
}

@inproceedings{vu2009differential,
  author    = {Duy Vu  and Aleksandra Slavkovic},
  title     = {Differential Privacy for Clinical Trial Data: Preliminary Evaluations},
  booktitle = {2009 IEEE International Conference on Data Mining Workshops},
  pages     = {138--143},
  year      = {2009},
  publisher = {IEEE}
}

@article{wang2018statistical,
  author  = {Wang, Yue and Kifer, Daniel and Lee, Jaewoo and Karwa, Vishesh},
  title   = {Statistical Approximating Distributions Under Differential Privacy},
  journal = {Journal of Privacy and Confidentiality},
  volume  = {8},
  number  = {1},
  year    = {2018}
}

@inproceedings{ferrando2022parametric,
  author    = {Ferrando, Cecilia and Wang, Shufan and Sheldon, Daniel},
  title     = {Parametric Bootstrap for Differentially Private Confidence Intervals},
  booktitle = {International Conference on Artificial Intelligence and Statistics},
  pages     = {1598--1618},
  year      = {2022},
  publisher = {PMLR}
}

@article{awan2025simulation,
  author  = {Awan, Jordan and Wang, Zhanyu},
  title   = {Simulation-Based, Finite-Sample Inference for Privatized Data},
  journal = {Journal of the American Statistical Association},
  volume  = {120},
  number  = {551},
  pages   = {1669--1682},
  year    = {2025}
}

@article{ju2022data,
  author  = {Nianqiao Ju and Jordan Awan and Ruobin Gong and Vinayak Rao},
  title   = {Data Augmentation {MCMC} for {B}ayesian Inference from Privatized Data},
  journal = {Advances in Neural Information Processing Systems},
  volume  = {35},
  pages   = {12732--12743},
  year    = {2022}
}

@article{gong2022exact,
  author  = {Ruobin Gong},
  title   = {Exact Inference with Approximate Computation for Differentially Private Data via Perturbations},
  journal = {Journal of Privacy and Confidentiality},
  volume  = {12},
  number  = {2},
  year    = {2022}
}

@inproceedings{kairouz2015composition,
  author    = {Kairouz, Peter and Oh, Sewoong and Viswanath, Pramod},
  title     = {The Composition Theorem for Differential Privacy},
  booktitle = {Proceedings of the 32nd International Conference on Machine Learning},
  series    = {Proceedings of Machine Learning Research},
  volume    = {37},
  pages     = {1376--1385},
  year      = {2015},
  editor    = {Bach, Francis and Blei, David},
  publisher = {PMLR}
}

@incollection{drechsler2021handbook,
  author    = {Drechsler, J{\"o}rg and Kifer, Daniel and Reiter, Jerome P. and Slavkovi{\'c}, Aleksandra},
  title     = {Handbook of Sharing Confidential Data},
  publisher = {Chapman and Hall/CRC},
  year      = {2024},
  edition   = {1},
  doi       = {10.1201/9781003185284}
}

@inproceedings{kulynych2024attackaware,
  author    = {Kulynych, Bogdan and Gomez, Juan Felipe and Kaissis, Georgios and Calmon, Flavio du Pin and Troncoso, Carmela},
  title     = {Attack-Aware Noise Calibration for Differential Privacy},
  booktitle = {Advances in Neural Information Processing Systems 37 (NeurIPS 2024)},
  pages     = {134868--134901},
  year      = {2024}
}
\bibliographystyle{apalike}

\appendix
\section{Proofs}
\subsection*{Theorem \ref{th: failure conditions}}
Assume that $f_\mu(0) < 1$, then
\begin{equation*}
\begin{split}
    p_{posterior} & = \frac{p_{\text{prior}}\{1 - f_\mu(0)\}}{(1 - p_{\text{prior}}) \times 0 + p_{\text{prior}} \{1 - f_\mu(0)\}} = 1 %, \\
    %& = \frac{p_{\text{prior}}\{1 - f_\mu(0)\}}{p_{\text{prior}} \{1 - f_\mu(0)\}}, \\
    %& = 1.
\end{split}
\end{equation*}
On the other hand, assume that $f_\mu$ is failing catastrophically, i.e., there exists $\alpha_0$ and $p_{prior}$ such that
\begin{align*}
\frac{p_{\text{prior}}\{1 - f_\mu(\alpha_0)\}}{(1 - p_{\text{prior}}) \times \alpha_0 + p_{\text{prior}} \{1 - f_\mu(\alpha_0)\}} = & 1 \\
p_{\text{prior}}\{1 - f_\mu(\alpha_0)\} = &(1 - p_{\text{prior}}) \times \alpha_0 + \\ & p_{\text{prior}} \{1 - f_\mu(\alpha_0)\} \\
(1 - p_{\text{prior}}) \times \alpha_0 = & 0,
\end{align*}
which implies that $\alpha_0 = 0$ as $p_{\text{prior}} > 1$. Then, we have
$$
\frac{p_{\text{prior}}\{1 - f_\mu(0)\}}{ p_{\text{prior}} \{1 - f_\mu(0)\}} = 1 ,
$$
which can be true only if $f_\mu(0) < 1$.$\square$

\subsection*{Theorem \ref{th: bounded risk}}
Following definition \ref{def: rel risk}, for all prior probabilities, the risk in bounded by $\{1 - f_\mu(\alpha)\} / \alpha$.

The later function is positive and continuous on $(0;1]$. So, if the trade-off function satisfies \eqref{eq: th1} by application of the extreme value theorem, it is bounded on $[0,1]$.
On the other hand, if the relative disclosure risk is bounded, then it is obvious that \eqref{eq: th1} hold true.

Finally, the trade-off function $f_\mu$ is lower bounded by $1 - \alpha$ excluding the possibility that $1 - f_\mu(\alpha)= o(\alpha)$. We thus conclude that \eqref{eq: th1} holds if and only if $1 - f_\mu(\alpha)$ is asymptotically equivalent to $ \rho\alpha$ with $\rho \geq 1$. $\square$.

\subsection*{Corollary \ref{cor: tail df}}
For a log-concave noise we have \citep{Dong2022}
$$
f_\mu(\alpha) = F\{F^{-1}(1 - \alpha) - \mu \}, 
$$
the relative disclosure risk is bounded by
\begin{align*}
    \frac{1 - f_\mu(\alpha)}{\alpha} & = \frac{1 - F\{F^{-1}(1 - \alpha) - \mu \}}{1 - F\{F^{-1}(1 - \alpha)\}} ,\\
    %& = \frac{1 - F\{F^{-1}(1 - \alpha) - \mu \}}{1 - F\{F^{-1}(1 - \alpha)\}} ,\\
    & = \frac{\bar{F}\{F^{-1}(1 - \alpha) - \mu \}}{\bar{F}\{F^{-1}(1 - \alpha)\}},
\end{align*}
which proves the corollary.  $\square$

\subsection*{Corollary \ref{cor: pure dp}}
 We have seen in section \ref{sec: f-dp} that the trade-off $f_\mu$ function of an $\epsilon$-DP mechanism $Q$ is upper bounded by equation \eqref{eq: tf pure dp} with $\delta = 0$. Thus for any $\alpha < (e^\epsilon - e^{-\epsilon}) / (1 - e^{\epsilon})$, we have $f_\mu(\alpha) \geq 1 - e^{\epsilon} \alpha$, which yields
$$
\frac{1 - f_\mu(\alpha)}{\alpha} \leq e^\epsilon,
$$
proving that the relative disclosure risk in upper bounded.$\square$

\end{document}